\documentclass[12pt,a4paper]{article}
\pdfoutput=1

\usepackage{amsmath,amssymb}
\usepackage[bookmarks=true]{hyperref}
\usepackage[nosort]{cite}
\usepackage[bulletsep]{collect}
\def\gfxon{\usepackage[final]{graphicx}}

\gfxon

\makeatletter
\let\old@startsection=\@startsection
\renewcommand{\@startsection}[6]{\old@startsection{#1}{#2}{#3}{#4}{#5}{#6\mathversion{bold}}}
\makeatother

\newcommand{\po}[1]{\frac{\partial}{\partial #1}}

\newcommand{\dpod}[1]{\partial_{#1}}

\newcommand{\corr}[1]{\langle #1 \rangle}

\newcommand{\eps}{\varepsilon}

\makeatletter \@addtoreset{equation}{section} \makeatother

\makeatletter
\let\old@makecaption=\@makecaption
\def\@makecaption{\small\old@makecaption}
\makeatother

\newcommand{\ham}{\mathcal{H}}

\newcommand{\gen}[1]{\mathfrak{#1}}

\newcommand{\moment}{\mathcal{P}}

\newcommand{\Bfield}{\mathcal{B}}

\newcommand{\vecs}{\mathcal{N}}
\newcommand{\const}{\mathcal{S}}

\newcommand{\ssN}{\mathcal{N}}

\renewcommand{\Re}{\mathop{\mathrm{Re}}}

\newcommand{\Nt}{\tilde N}

\makeatletter
\newcommand{\ellSN}{\mathop{\operator@font sn}\nolimits}
\newcommand{\ellCN}{\mathop{\operator@font cn}\nolimits}
\newcommand{\ellDN}{\mathop{\operator@font dn}\nolimits}
\newcommand{\ellAM}{\mathop{\operator@font am}\nolimits}
\newcommand{\ellK}{\mathop{\smash{\operator@font K}\vphantom{a}}\nolimits}
\newcommand{\ellE}{\mathop{\smash{\operator@font E}\vphantom{a}}\nolimits}
\makeatother

\ifx\genfrac\sdflkaj

\else

\fi
\newcommand{\sfrac}[2]{{\textstyle\frac{#1}{#2}}}
\newcommand{\half}{\sfrac{1}{2}}

\newcommand{\acomm}[2]{\{#1,#2\}}

\newcommand{\covder}{\mathcal{D}}

\newcommand{\nln}{\nonumber\\}
\newcommand{\nl}[1][0pt]{\nonumber\\[#1]&\hspace{-4\arraycolsep}&\mathord{}}

\newcommand{\earel}[1]{\mathrel{}&\hspace{-2\arraycolsep}#1\hspace{-2\arraycolsep}&\mathrel{}}
\newcommand{\eq}{\earel{=}}
\newcommand{\beq}{\begin{equation}}
\newcommand{\eeq}{\end{equation}}

\def\[{\begin{equation}}
\def\]{\end{equation}}
\def\<{\begin{eqnarray}}
\def\>{\end{eqnarray}}

\def\a{\alpha}
\def\b{\beta}

\def\eps{\varepsilon}
\def\g{\gamma}

\def\m{\mu}
\def\p{\partial}
\def\s{\sigma}
\def\t{\theta}
\def\z{\zeta}

\def\f{\frac}
\def\be{\begin{equation}}
\def\ee{\end{equation}}
\def\bea{\begin{eqnarray}}
\def\eea{\end{eqnarray}}
\def\ba{\begin{array}}
\def\ea{\end{array}}
\def\bc{\begin{center}}
\def\ec{\end{center}}
\def\bl{\begin{flushleft}}
\def\el{\end{flushleft}}
\def\br{\begin{flushright}}
\def\er{\end{flushright}}
\def\l{\left}
\def\r{\right}

\makeatletter
\def\mr@ignsp#1 {\ifx\:#1\@empty\else #1\expandafter\mr@ignsp\fi}%
\newcommand{\multiref}[1]{\begingroup
\xdef\mr@no@sparg{\expandafter\mr@ignsp#1 \: }%
\def\mr@comma{}%
\@for\mr@refs:=\mr@no@sparg\do{\mr@comma\def\mr@comma{,}\ref{\mr@refs}}%
\endgroup}
\makeatother

\newcommand{\hypref}[2]{\ifx\href\asklfhas #2\else\href{#1}{#2}\fi}

\newcommand{\secref}[1]{Sec.~\multiref{#1}}

\newcommand{\figref}[1]{Fig.~\multiref{#1}}
\renewcommand{\eqref}[1]{(\multiref{#1})}
\newcommand{\lagr}{\mathcal{L}}

\def\chfin#1{{#1}}

\newcommand{\Integers}{\mathbb{Z}}

\newcommand{\Complex}{\mathbb{C}}

\newcommand{\CP}[1]{\mathbb{CP}^{#1}}
\newcommand{\WCP}[1]{\mathbb{WCP}^{#1}}

\ifx\href\asklfhas\newcommand{\href}[2]{#2}\fi
\newcommand{\arxivno}[1]{\href{http://arxiv.org/abs/#1}{#1}}

\begin{document}

\begin{flushright}\footnotesize
\texttt{ArXiv:\arxivno{1009.6207}}\\
\texttt{FTPI-MINN-10/26} \\
\texttt{UMN-TH-2918/10}
\vspace{0.5cm}
\end{flushright}
\vspace{0.3cm}

\renewcommand{\thefootnote}{\arabic{footnote}}
\setcounter{footnote}{0}
\begin{center}%
{\Large\textbf{\mathversion{bold}
Large-$N$ Solution of the Heterotic Weighted Non-linear Sigma-Model}
\par}

\vspace{1cm}%

\textsc{Peter Koroteev, Alexander Monin and Walter Vinci}

\vspace{5mm}

\textit{University of Minnesota, School of Physics and Astronomy\\%
116 Church Street S.E. Minneapolis, MN 55455, USA}

\vspace{7mm}

\thispagestyle{empty}

\texttt{koroteev, monin, vinci@physics.umn.edu} \\

\par\vspace{1cm}

\vfill

\textbf{Abstract}\vspace{5mm}

\begin{minipage}{12.7cm}
We study a heterotic two-dimensional $\ssN=(0,2)$ gauged non-linear sigma-model whose target space is a weighted complex projective space. We consider the case with $N$ positively  and $\Nt=N_{F}-N$ negatively charged fields. This model is \chfin{believed to give a} description of the low-energy physics of a non-Abelian semi-local vortex in a four-dimensional $\ssN=2$ supersymmetric $U(N)$ gauge theory with $N_{F}>N$ matter hypermultiplets. The supersymmetry in the latter theory  is broken down to $\ssN=1$ by a mass term for the adjoint fields. We solve the model in the large-$N$ approximation and explore a two-dimensional subset of the mass parameter space for which a discrete $\mathbb Z_{N-\Nt}$ symmetry is preserved. Supersymmetry is generically broken, but it is preserved for special values of the masses where a new branch opens up and the model becomes super-conformal.

\end{minipage}

\vspace{3mm}

\vspace*{\fill}

\end{center}

\newpage

\section{Introduction}\label{Sec:Intro}

For many years two-dimensional $\mathbb{CP}^{N-1}$  sigma-models have been providing extremely useful insights into the physics of four-dimensional non-Abelian gauge theories. One of the most important features the two types of theories share is the non-perturbative generation of a mass gap \cite{Witten:1978bc,Gorsky:2005ac}.

The connection has been tightened up thanks to recent results in gauge theories with extended supersymmetry. First it has been proven that the $\ssN=(2,2)$ extension of the two-dimensional $\CP{N-1}$ sigma-model has the same spectrum of massive BPS states as the $\ssN=2$ four-dimensional $SU(N)$ gauge theory with $N$ hypermultiplets, provided that the parameters of the two theories are identified in a proper way \cite{Dorey:1998yh, Dorey:1999zk}. Remarkably, the correspondence holds at both the classical and quantum levels. The physical reason behind that was unclear until it was realized that the correct two-dimensional model arises naturally as an effective theory on string-like solitons existing in the four-dimensional bulk theory. The key point was the discovery of non-Abelian vortices \cite{Hanany:2003hp,Auzzi:2003fs,Gorsky:2004ad,Eto:2005yh}, which posses internal degrees of freedom with non-trivial dynamics. The fluctuations of the fields around the vortex configuration can be thought of as the original particles confined to the world-sheet of the vortex, due to the Higgs screening \cite{Hanany:2004ea,Shifman:2004dr}.

In an attempt to \chfin{further study} the relationship between the theories (in a set-up which may be closer to the real QCD) one introduces mass terms which decouple the adjoint scalar fields \cite{Gorsky:2007ip} present in $\ssN=2$ theories. Having done that, one breaks supersymmetry down to $\ssN=1$. SUSY breaking terms correspond to a very interesting deformation of the vortex world-sheet theory which gives rise to a particular type of $\ssN=(0,2)$ $\CP{N-1}$ sigma-model called ``heterotic'' \cite{Edalati:2007vk,Shifman:2008wv,Bolokhov:2009wv}.

The heterotic $\mathbb{CP}^{N-1}$ sigma-model was first analyzed in Ref. \cite{Tong:2007qj}, then it was solved in the large-$N$ approximation in Refs. \cite{Shifman:2008kj,Bolokhov:2010hv}. The model shows a rich set of phenomena like spontaneous supersymmetry breaking and transitions between Higgs and Coulomb/confining phases. Again, the two-dimensional sigma-models have proven to capture important properties of the corresponding four-dimensional bulk theories \cite{Tong:2007qj,Bolognesi:2009ye}.

In this paper we consider a particular extension of the  $\CP{N-1}$ sigma-model which can be obtained by gauging $N$ positively charged fields. Considering additional $\Nt=N_{F}-N$ matter multiplets with negative charge, we obtain what is called a  ``weighted'' $\CP{N_{F}-1}$ (or $\WCP{N_{F}-1}$) sigma-model\footnote{The notation is borrowed from a previous work of one of the authors \cite{Eto:2007yv}, where it was used in connection with the moduli space of semi-local vortices. ln the context of  algebraic geometry, where these spaces are well studied, they are more correctly referred to as $\mathcal O(-1)^{\Nt}$ line bundles over $\CP{N-1}$.}. The target space $\WCP{N_{F}-1}$ contains $\mathbb{CP}^{N-1}$ as a subspace. The crucial point is that the weighted projective space is not compact. The model was proposed in Ref. \cite{Hanany:2003hp} as the low-energy description of non-Abelian semi-local vortices. Semi-local vortices appear in gauge theories when large global symmetries are present \cite{Preskill:1992bf}. These symmetries are usually realized as flavor symmetries by introducing additional matter fields. The main feature of these vortices is the existence of a new set of degenerate solutions with arbitrary size \cite{Achucarro:1999it,Shifman:2006kd,Auzzi:2008wm}. In fact, this property makes semi-local vortices quite similar to instantons and lumps \cite{Hindmarsh:1991jq,Eto:2007yv,Collie:2009iz,Eto:2009bz}. Employing a D-brane construction, the authors of Ref. \cite{Edalati:2007vk}  found the unique heterotic deformation to this model which could arise when the symmetry breaking term in the bulk theory is turned on. Motivated by this,  we use the large-$N$ techniques exploited in Refs. \cite{Witten:1978bc,Shifman:2008kj,Bolokhov:2010hv} to solve the model and understand its physics.

\chfin{In the} correspondence between the supersymmetric QCD and the $\ssN=(2,2)$ sigma-model mentioned above, the complex masses of the hypermultiplets in the former theory coincide with the \textit{twisted} masses of the latter theory. In the model addressed in the current paper, we introduce $N$ twisted masses $m_{i}$ for each positively charged field and additional $\Nt$ twisted masses $\mu_{j}$ for each negatively charged field. As is known from the $\CP{N-1}$ sigma-model, the values of the twisted masses control the phases of the theory. Indeed, if the masses are much bigger than the dynamically generated scale $\Lambda$, the theory is essentially classical, whereas quantum effects become significant for $m_{i},\mu_{j}\lesssim \Lambda$. Due to larger variety of twisted masses, the phase diagram of the theory is quite complicated. We shall consider a particular choice of the masses which preserves a discrete symmetry, by appropriately putting them on two circles of radii $m$ and $\mu$. We thus focus on the determination of the phase diagram of the model in terms of these two parameters.

The supersymmetric  $\CP{N-1}$ sigma-model is known to have an exact ``twisted'' superpotential \cite{D'Adda:1982eh, Cecotti:1992rm, Dorey:1998yh} which is similar to the Veneziano--Yankielowicz superpotential \cite{Veneziano:1982ah}.  It can be straightforwardly generalized to the weighted sigma-model \cite{Cecotti:1992rm, Hanany:1997vm, Dorey:1999zk}. The exact superpotential depends only on the twisted chiral superfield containing the gauge multiplet and twisted masses of the theory. Once the superpotential is known one can in principle determine the full  BPS spectrum, including  the vacua of the theory for any $N$ and $\tilde N$. However, if we break half of the supersymmetries by introducing the heterotic deformation, we cannot rely on the existence of an exact superpotential anymore, and we have to dwell on a more robust technique of solving quantum theories at strong coupling, like the large-$N$ approximation. 

The paper is organized as follows. In \secref{Sec:SuperfieldCPN} we introduce the model and discuss its quantum aspects. In \secref{Sec:Large-NSolution} we present the master set of equations which gives the vacuum expectation values of all the fields. We solve them exactly in the massless case, while we give approximate analytical solutions and numerical evaluations in various regimes for non-zero masses. In \secref{Sec:Spectrum} we discuss the one-loop low-energy effective action which describes excitations above the vacua found earlier. \secref{Sec:Discussion} contains conclusions and discussions. 
Various technical details and useful formulae are given in the appendices.

\section{The Model}\label{Sec:SuperfieldCPN}

In this part of the paper we formulate the sigma-model in gauged approach and discuss its moduli space.
First we consider the $(2,2)$ sypersymmetric model and then introduce the heterotic deformation.

\subsection{$\ssN=(2,2)$ weighted non-linear sigma-model} 

Let us start by introducing the undeformed  $\ssN=(2,2)$  weighted sigma-model\footnote{Many gauged sigma-models which are studied in the literature, including this one, follow from a very generic construction developed by Distler and Kachru \cite{Distler:1993mk}.} $\WCP{N_{F}-1}$. A detailed discussion of these models can be found in Ref. \cite{Witten:1993yc}, where a relationship with Landau-Ginzsburg models is considered. The same models can be studied in the mirror representation \cite{Coleman:1974bu,Fateev:1978xn,Hori:2000kt}. The model can be built out of $N$ positively charged fields $n_{i}$, $\Nt$ negatively charged fields $\rho_{j}$ and a non-dynamical auxiliary field. The full Lagrangian, including the fermionic superpartners can be written in a superfield formalism which make supersymmetry manifest (see \secref{Sec:SuperField}). The Lagrangian \eqref{eq:N1Heterotic} has the following  component expansion 
\bea
\mathcal {L}_{\WCP{N_F-1}} & = & \l |\nabla _ \m  n _ i \r | ^ 2 + \l |\nabla _ \m  \rho _ j \r | ^ 2 - | \s | ^ 2 |n _ i| ^ 2 - | \s | ^ 2 |\rho_j| ^ 2
- D \l ( |n _ i | ^ 2 - |\rho _ j| ^ 2 -  r _ 0 \r )  \nonumber \\
\nonumber \\
 & + &  i \bar \xi_{L,\, i} \nabla _ R \xi _ L ^ i
+ i \bar \xi_{R,\, i} \nabla _ L \xi _ R ^ i + i \bar \eta_{L,\, j} \nabla _ R \eta _ L ^ j
+ i \bar \eta_{R,\, j} \nabla _ L \eta _ R ^ j + \nonumber \\ \nonumber \\
& + & \left[ i\bar n_i \l( \lambda _ L \xi _ R ^ i- \lambda _ R \xi _ L ^ i \r )  - i  \s \bar \xi_{R,\, i} \xi _ L ^ i  - i \bar \rho _ j \l ( \lambda _ L \eta _ R ^ j - \lambda _ R \eta _ L ^ j \r )  + i \s \bar \eta _ R ^ j \eta _ L ^ j + \text {H.c.} \r ] \,, \nonumber \\
\label{eq:complagr} 
\eea
where the covariant derivatives are given by
\[
\nabla_{\mu} n_{i}=(\partial_{\mu}-iA_{\mu})n_{i}, \quad \nabla_{\mu} \rho_{j}=(\partial_{\mu}+iA_{\mu})\rho_{j}\,.
\label{eq:CovDer}
\]
The fields $A_{\mu}$, $\sigma$, $\lambda_{L,R}$ and $D$ all belong to the same $\ssN=2$ supermultiplet, they are non-dynamical, and can  be integrated out using their equations of motion. However, as we shall see later, in strongly coupled phases these auxiliary fields do become dynamical and describe particles in the low energy effective theory.

The model has a unique parameter which determines the strength of the interactions, the two-dimensional Fayet-Iliopoulos term $r_{0}$ \cite{Fayet:1974jb}. Classically, the model has a continuous set of vacua determined by the vacuum equation
\[
\sum\limits_{i=0}^{N-1} |n_i | ^ 2 - \sum\limits_{j=0}^{\tilde N -1} |\rho_ j| ^ 2 =  r_0\,.
\]
The first and the most important quantum effect is the generation of a dynamical scale $\Lambda$ through dimensional transmutation. In fact, the Fayet-Iliopoulos term gets renormalized, flowing with respect to the energy scale $\epsilon$ through the following one loop expressions
\[\label{eq:reps}
r(\epsilon)=r_{0}-\frac{N-\Nt}{4\pi}\log\left( \frac{M_{UV}^{2}}{\epsilon^{2}}\right)\equiv-\frac{N-\Nt}{4\pi}\log\left(\frac{\Lambda^{2}}{\epsilon^{2}}\right).
\]
The theory is thus asymptotically free for $N>\Nt$. From the expression above we can also guess that for $N=\Nt$ we have super-conformal theory, and this is indeed the case \cite{Witten:1993yc}. 

Actually, thanks to supersymmetry, \eqref{eq:reps} is exact in perturbation theory because of  the vanishing of higher order contributions. Furthermore, integrating out the matter fields in the functional integral we can find an exact superpotential for the field $\sigma$  \cite{Witten:1993yc, D'Adda:1982eh, Cecotti:1992rm, Hanany:1997vm}
\<
W(\sigma)&=& \frac{N-\Nt}{4\pi}\sigma\left(\log\left(\frac{\sigma}{\Lambda}\right)-1\right).
\label{eq:masslesssuperpot}
\>
This superpotential includes all the non-perturbative instantonic contributions to the functional integral.
At the classical level the theory has two $U(1)$ R-symmetries, $U(1)_{R}\times U(1)_{V}$. The first one is an axial symmetry, under which $\sigma$ has charge +2. This symmetry is anomalous and is broken down to $\mathbb Z_{2N-2\Nt}$ by the one-loop corrections. By minimization of the superpotential \eqref{eq:masslesssuperpot} we find $N-\Nt$ massive vacua. We will discuss in more details the vacuum structure of the theory  in \secref{Sec:Large-NSolution}.

\subsection{$\ssN=(0,2)$ weighted sigma-model: heterotic deformation} 

As is well-known from early studies of two-dimensional supersymmetric sigma-models \cite{Zumino:1979et}, there is no smooth $\ssN=(0,2)$ deformation of the $\ssN=(2,2)$ $\CP{N-1}$ sigma-model\footnote{See Refs. \cite{Edalati:2007vk,Shifman:2005st} for a discussion of this issue in a context related to non-Abelian vortices}. On the other hand, it is possible to have deformation of the $\Complex\times\CP{N-1}$ model, which is the relevant effective theory emerging in when studying the non-Abelian vortices (the $\Complex$ factor describes the translation modes of the vortex). From the additional $\mathbb C$ piece, one can keep only a right-handed fermion, while the scalar and left-handed fermionic super-partners is free. A similar situation occurs for the weighted sigma-model\footnote{In fact, it is possible to introduce $\ssN=(0,2)$ deformations of the weighted sigma-model without introducing any new degrees of freedom, or $\Complex$ factors. However, all the possible deformations different from the one considered in the text do not arise in the context of non-Abelian vortices. Nevertheless, it may be interesting to study the effects of such deformations. For more details on this aspect, see Ref. \cite{Edalati:2007vk}.}. As a result we consider the following Lagrangian
\bea
\mathcal {L}_{\WCP{N_F-1}}^{het} & = & \mathcal {L}_{\WCP{N_F-1}}+ \sfrac{i} {2} \bar \z _ R \p _ L \z _ R- 2 |\omega| ^ 2 |\s| ^ 2  - \l [ i \omega \lambda _ L \z _ R + \text {H.c.} \right]\,.
\eea
The heterotic coupling $\omega$ is introduced by means of an additional right-handed fermion $\zeta_{R}$. Obviously the modification dramatically changes the physics of the sigma-model at hand. For example, the Witten index is modified from $N-\Nt$ to zero as in the  $\CP{N-1}$ case.  This observation is indeed consistent with supersymmetry breaking \cite{Shifman:2008kj,Witten:1982df} occurring in the model.

\paragraph{Adding the twisted masses.}

Twisted masses can be easily introduced into the model by first gauging the $U(1)^{N_{F}-1}$ independent flavor symmetries and then setting to zero all the fields in the additional twisted multiplets but not the lowest components \cite{Hanany:1997vm}. The resulting Lagrangian takes the following form
\<\label{eq:WCPNTwistedMasses}
\lagr^{het}_{\WCP{N_F-1}} \eq \left|\nabla_\mu  n_i \right|^2 + \l |\nabla _ \m  \rho _ j \r | ^ 2 
+ i \bar \xi_{L,\, i} \nabla_R \xi_L^i + i\bar \xi_{R,\, i} \nabla_L \xi_R^i + i\bar \eta_{L,\, j} \nabla_R \eta_L^j + i \bar \eta_{R,\, j} \nabla_L \eta_R^j\nl 
-  \sum\limits_{i=0}^{N-1} \left|\sigma- m_i\right|^2 |n_i| ^ 2 - \sum\limits_{j=0}^{\tilde N-1} \left|\sigma- \mu_j\right|^2 |\rho_j| ^ 2 - D \left(|n_i|^2- |\rho_j|^2 -  r_0\right)  \nl
+ \l[ i \bar n _ i \l ( \lambda _ L \xi _ R ^ i- \lambda _ R \xi _ L ^ i \r )  - i \sum\limits_{i=0}^{N-1} \left(\sigma- m_i\right) \bar \xi_{R,\, i} \xi_L^i + \text {H.c.} \r ]  \nl
+ \l[ - i  \bar \rho _ j \l ( \lambda _ L \eta _ R ^ j - \lambda _ R \eta _ L ^ j \r )  + i \sum\limits_{j=0}^{\tilde N-1} \left({\sigma}- \mu_j\right)\bar \eta_{R,\,j}\eta_L^j + \text {H.c.} \r ]  \nl
+ \sfrac{i} {2} \bar \z _ R \p _ L \z _ R - \l [ i  \omega \lambda _ L \z _ R + \text {H.c.} \right]- 2 |\omega|^2|\sigma|^2\,.
\>
For zero values of the twisted masses there is a $U(1)$ R-symmetry under which the fermions $\xi_R^i, \eta_R^j,\lambda_R\,\, (\xi_L^i, \eta_L^j,\lambda_L )$  have charge $+1 (-1)$, whereas $\sigma$ has charge $+2$. A generic choice of the masses $m_{i}$ and $\mu_{j}$ breaks this symmetry completely. Instead, we make the following choice for the masses
\<\label{eq:CircleMasses}
m_k  \eq   m \,e^{2\pi i\frac{k}{N}}\,,\quad k = 0,\dots, N-1\,,\nln
\mu_l  \eq   \mu\, e^{2\pi i\frac{l}{\tilde N}}\,,\quad l = 0,\dots, \tilde N-1\,.
\>
For the further convenience we define a new constant $\alpha=\Nt/N$. Notice that in the $N \to \infty$ limit, the masses are distributed uniformly on circles with radii $|m|$ and $|\mu|$ correspondingly. We consider $m$ and $\mu$ to be real. There are  particular choices of $\alpha$ which are interesting because they leave some residual discrete symmetry on the classical level. In particular, if $N$ and $\tilde N$ have $N-\Nt$ as a common divisor, a discrete  $\mathbb Z_{N-\Nt}$ symmetry is preserved\footnote{This symmetry is a combination of the flavor and R symmetry.}. As we shall later see in \secref{Sec:Large-NSolution}, in quantum theory VEV of $\sigma$ breaks this symmetry, however, for certain values of the twisted masses \eqref{eq:CircleMasses} $\corr{\sigma}=0$ and the symmetry gets restored.

\section{Large-$N$ Solution}\label{Sec:Large-NSolution}

In this section we solve the model in the large-$N$ approximation, closely following the analysis of Refs. \cite{Shifman:2008kj,Bolokhov:2010hv}. Since the $n^i\,, \rho^j\,, \xi^i\,, \eta^j$ fields appear in the action quadratically, we can perform the Gaussian integration over these fields. We integrate over all but the following four fields $(n^{0}, \rho^{0}, \xi^{0}, \eta^{0})$. The scalar fields $(n^{0}, \rho^{0})$ will represent the helpful set of the order parameters defining various phases of the theory.

The Gaussian integration leads to the following determinants
\<
\nonumber \\
\prod_{i=1}^{N-1}\left[  \frac{\det\left((\partial_{k}+i A_{k})^{2}+D+|\sigma-m_{i}|^{2}\right)}{  \det\left((\partial_{k}+i A_{k})^{2}+|\sigma-m_{i}|^{2}\right)} \right]\prod_{j=1}^{\Nt-1}\left[  \frac{\det\left((\partial_{k}-i A_{k})^{2}-D+|\sigma-\mu_{j}|^{2}\right)}{  \det\left((\partial_{k}-i A_{k})^{2}+|\sigma-\mu_{j}|^{2}\right)} \right]\,.
\nonumber \\
\label{eq:dets}
\>
The large-$N$ approximation is technically equivalent to a one-loop calculation of the above determinants, where we can also drop the gauge fields \cite{Witten:1978bc}. The result gives an effective potential for the $\sigma$ field\footnote{For a discussion of the relationship between the Large-$N$ potential and the exact $\ssN=(2,2)$ superpotential \eqref{eq:masslesssuperpot} see Ref. \cite{Bolokhov:2010hv}. It is indeed possible to reconstruct a full exact potential like \eqref{eq:masslesssuperpot} from this expression, by noticing that the large-$N$ expression must give, at the first  linear order in $D$, the following term: $D (W'(\sigma)+h.c)$. We thank A. Vainshtein for this observation. } 
\<
V_{1-loop}\eq\frac{1}{4\pi}\sum\limits_{i=1}^{N-1}\left(-\left(D+\left|\sigma- m_i\right|^2\right)\log\frac{\left|\sigma- m_i\right|^2+D}{\Lambda^2}+\left|\sigma- m_i\right|^2\log\frac{\left|\sigma- m_i\right|^2}{\Lambda^2}\right)\nl
-\frac{1}{4\pi}\sum\limits_{j=1}^{\tilde N-1}\left(-\left(D-\left|\sigma- \mu_j\right|^2\right)\log\frac{\left|\sigma- \mu_j\right|^2-D}{\Lambda^2}-\left|\sigma- \mu_j\right|^2\log\frac{\left|\sigma- \mu_j\right|^2}{\Lambda^2}\right)\nl
+\frac{N-\tilde N}{4\pi}D\,.
\label{eq:EffectivePotentialoneloop}
\>
To get the above result we have again traded the UV cut-off for the scale $\Lambda$.
Including the pieces already present at the classical level we get the expression for the effective potential
\<\label{eq:EffectivePotentialFull}
V_{eff}\eq V_{1-loop}+\left(\left|\sigma- m_0\right|^2+D\right)|n_0|^2 +\left(\left|\sigma- \mu_0\right|^2-D\right)|\rho_0|^2+\frac{uN}{4\pi}\,|\sigma|^2 \,,
\>
where we set $u=8\pi |\omega|^2/N$.

\paragraph{Vacuum equations.} 

The extremization\footnote{The solution of the vacuum equations for $D$ gives $V_{eff}$ a maximum rather than a minimum. This fact, being usual in supersymmetric gauge theories, is consistent since the $D$ field is not dynamical. We get a true minimum with respect to the $\sigma$ field. } of this potential with respect to $n_0$ and $\rho_0$, $D$ and $\sigma$ gives us the master set of equations which determines the vacuum structure of the theory
\begin{align}\label{eq:VacEqDsigma}
& \left(\left|\sigma- m_0\right|^2+D\right)n_0=0\,,\quad \left(\left|\sigma- \mu_0\right|^2-D\right)\rho_0=0\,, \\
&\frac{1}{4\pi}\sum\limits_{i=1}^{N-1}\log\frac{\left|\sigma- m_i\right|^2+D}{\Lambda^2} -\frac{1}{4\pi}
\sum\limits_{j=1}^{\tilde N-1} \log\frac{\left|\sigma- \mu_j\right|^2-D}{\Lambda^2}=|n_0|^2-|\rho_0|^2 \,,\notag\\
& \frac{1}{4\pi}\sum\limits_{i=1}^{N-1} \left({\sigma}- m_i\right)\log\frac{\left|\sigma- m_i\right|^2+D}{\left|\sigma- m_i\right|^2}   +\frac{1}{4\pi}\sum\limits_{j=1}^{\tilde N-1}  \left({\sigma}- \mu_j\right)\log\frac{\left|\sigma- \mu_j\right|^2-D}{\left|\sigma- \mu_j\right|^2} =\notag\\ & = \left({\sigma}- m_0\right)|n_0|^2+\left({\sigma}- \mu_0\right)|\rho_0|^2  +\frac{uN}{4\pi}\,\sigma\,.
\end{align}
The second equation above gives us the renormalized coupling constant
\[
r= |n_0|^2 - |\rho_0|^2\,.
\label{eq:rencoupling}
\]
In the next section we shall solve the weighted heterotic $\mathbb{CP}^{N-1}$ model in the large-$N$ approximation. First we address the massless case, and then work out the more involved model with twisted masses.

\subsection{Massless case}

Let us warm-up with the problem when all twisted mass are zero. We will be able to investigate more easily all the features which will be also present in the massive case. The potential \eqref{eq:EffectivePotentialFull} takes much simpler form now\footnote{Notice that in this case we have integrated out all the fields.}
\<
V_{eff}\eq\frac{N}{4\pi}\left(D\log\frac{\Lambda^2}{|\sigma|^2+D}+|\sigma|^2\log\frac{|\sigma^2|}{|\sigma|^2+D}\right)\nl
-\frac{\tilde N}{4\pi}\left(D\log\frac{\Lambda^2}{|\sigma|^2-D}-|\sigma|^2\log\frac{|\sigma^2|}{|\sigma|^2-D}\right)\nl
+\frac{N-\tilde N}{4\pi}D+\frac{uN}{4\pi}|\sigma|^2\,,
\label{eq:EffectivePotentialMassless}
\>
from which  the corresponding vacuum equations follow
\<\label{eq:VacuaEqns}
\log\frac{|\sigma|^2+D}{\Lambda^2}-\alpha\log\frac{|\sigma|^2-D}{\Lambda^2} \eq 0\,,\nln
\sigma\log\left(1+\frac{D}{|\sigma|^2}\right)+\sigma\alpha\log\left(1-\frac{D}{|\sigma|^2}\right) \eq u\sigma\,.
\>
Let us rewrite them in a more compact form
\<\label{eq:VacuaEqnsxs}
(1+x)(1-x)^\alpha \eq e^u\,,\nln
(1-x)^{\frac{\alpha }{1-\alpha }} (1+x)^{\frac{1}{\alpha-1}} \eq s\,,
\>
where we introduced the following dimensionless parameters
\[
s\equiv |\sigma|^2/\Lambda^2,\, \quad x \equiv  D/|\sigma|^2\,.
\]

Let us first discuss the undeformed case ($u=0$). From the first equation of \eqref{eq:VacuaEqnsxs} we get $x=0$, and from the second $s=1$, thus
\[
|\sigma_{\Lambda}|=\Lambda, \quad D=0\,.
\label{eq:lambdavaccmassless}
\]
The vanishing of the VEV of $D$ implies unbroken supersymmetry. The VEV of $\sigma$, on the contrary, lies on a circle\footnote{This is a natural result if we keep only the leading terms in the large-$N$ approximation. Separating vacua into a discrete set should be possible, in principle, by considering sub-leading corrections to the potential.}. We can compare this result with the exact $\ssN=(2,2)$ solution at finite-$N$ by minimizing the potential \eqref{eq:masslesssuperpot}, { from which we get the vacuum equation
\[
\sigma^{N-\tilde N} = \Lambda^{N-\tilde N}\,.
\label{eq:exactvacuamassless}
\]
}  There are $N-\Nt$ vacua characterized by the vacuum  expectation value of $\sigma$
\[
\sigma_{\Lambda,k} = \Lambda e^{2\pi i\frac{k}{N-\tilde N}}\,.
\label{eq:Lambdavac}
\]
We can see that in the large-$N_F$ limit the number of vacua becomes infinite and uniformly distributed on the circle.

\begin{figure}[!h]
\begin{center}
\includegraphics[height=5cm, width=7.7cm]{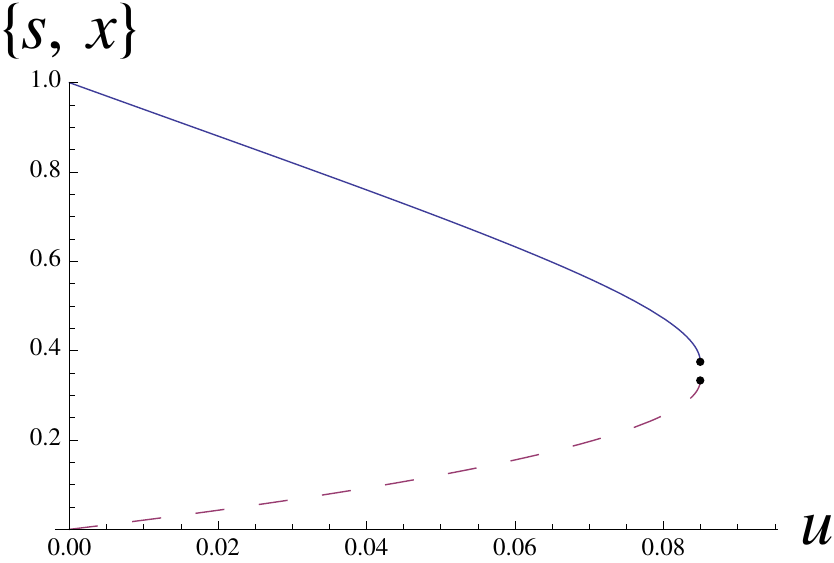}  \quad
\includegraphics[height=5cm, width=7.7cm]{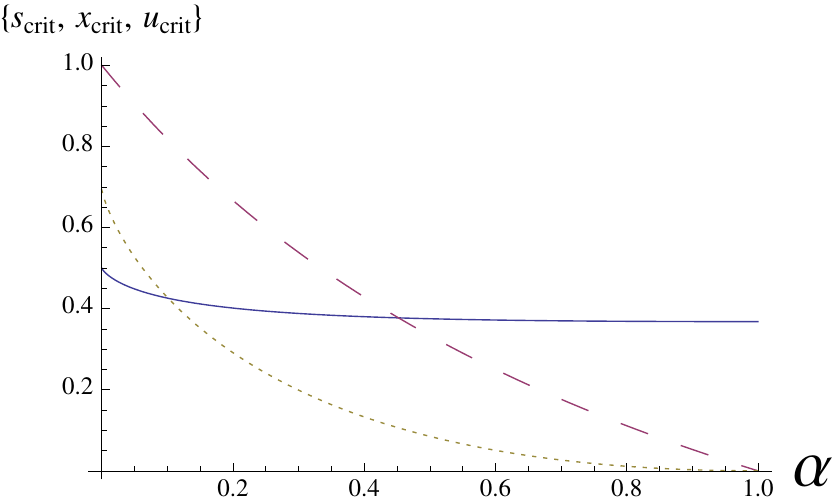}
\caption{On the left plot, values of $s$ and $x$ are shown as a function of $u$, for $\alpha=0.5$. On the right plot, the critical values as functions of $\alpha$ are shown. $s$ is solid-blue, $x$  dashed-red and  $u$  dotted-yellow.}
\label{fig:UDomainM0}
\end{center}
\end{figure}

Let us now turn on the heterotic deformation. The first equation from \eqref{eq:VacuaEqnsxs} gives us $x$, the second one can be used to find both $D$ and $\sigma$ in the vacuum configuration.  The r.h.s of the first equation  has an upper bound. There is thus a critical value $u_{crit}$ for the heterotic deformation such that there are no solutions for larger $u$. Maximization of this term gives
\[\label{eq:UcritScrit}
u_{crit}=\log\left[\frac{2(2\alpha)^\alpha}{(1+\alpha)^{\alpha+1}}\right] \quad {\rm at} \quad x_{crit} = \frac{1-\alpha}{1+\alpha},\quad s_{crit}=\frac12(1+\alpha)\alpha^{\frac{\alpha}{1-\alpha}}\,.
\]
The numerical solution of equations \eqref{eq:VacuaEqnsxs} is presented in \figref{fig:UDomainM0}.
Note that $x$ is always smaller than unity. This is consistent with the fact that larger values of $D$  ($D>|\sigma^{2}|$) would imply imaginary masses for the scalar particles, as it can easily be seen from \eqref{eq:complagr}. 

\begin{figure}[ht]
\begin{center}
\includegraphics[height=4.5cm, width=7cm]{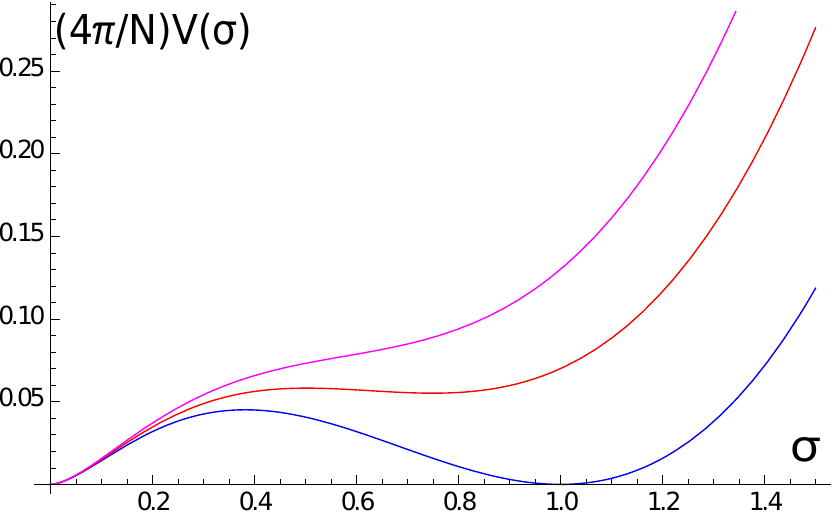} \quad \includegraphics[height=4.5cm, width=7cm]{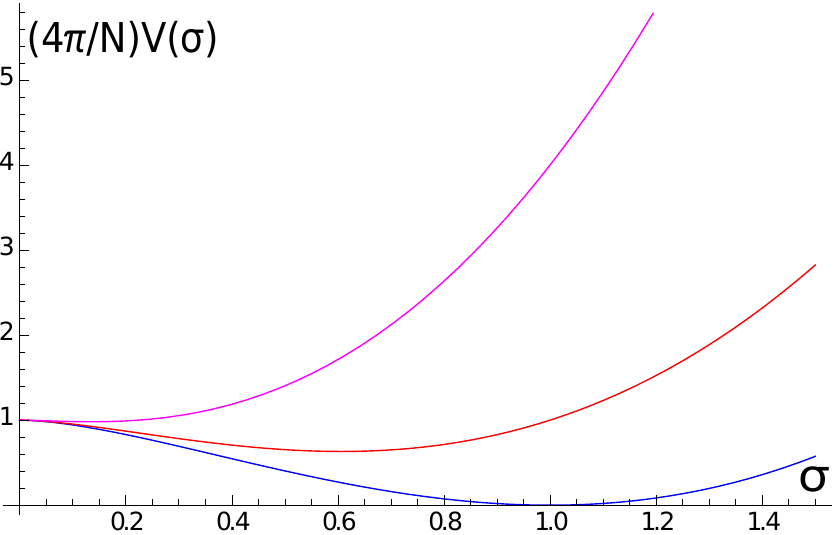}\\
\caption{On the left: one-loop effective potential for the weighted $\WCP{N_{F}-1}$ sigma-model, $u=0, 0.07, 0.13$ (from the lowest to the highest curve) and $\alpha=0.5$ as functions of $\sigma$ in units of $\Lambda$. The lower lying plot corresponds to the unbroken SUSY ($u=0$) where the vacuum energy of both zero vacuum and $\Lambda$ vacuum is equal to zero. When we enhance the heterotic deformation the vacuum energy of the $\Lambda$ vacuum becomes nonzero (the vacuum becomes metastable), whereas it always vanishes for the zero vacuum. At some value of the deformation parameter, $u_{crit}$ the metastable vacuum ceases to exist. On the right: potential for the ordinary $\CP{N-1}$ sigma-model, $u=0,1,4$. The vacuum value of $\sigma$ approaches zero for large $u$, but there is no loos of vacua, as soon as the deformation is kept finite. }
\label{fig:PotentialsM0}
\end{center}
\end{figure}

The disappearance of the solutions which minimize the energy becomes clear after we look at the plots of the effective potential at different values of the heterotic deformation parameter. Using \eqref{eq:VacuaEqns} we can find the auxiliary field $D$  and substitute it into the effective potential \eqref{eq:EffectivePotentialMassless}, which we can now plot in 
\figref{fig:PotentialsM0} as a function of $\sigma$.

\begin{figure}[ht]
\begin{center}
\quad \includegraphics[height=5cm, width=7cm]{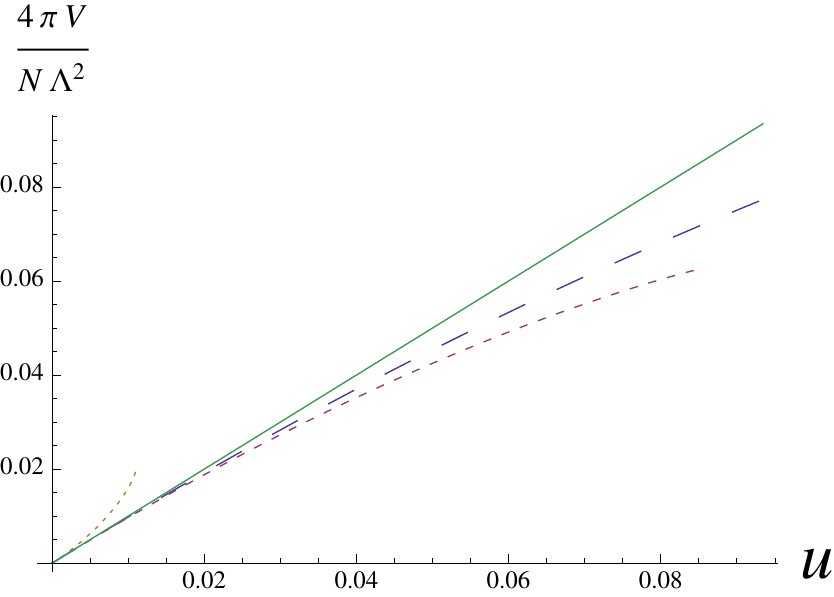}  
\caption{Numerical solutions for the vacuum energy of the massive vacua for various values of $\alpha=.2,.5,.8$, from the lowest to the highest curve. The solid line is the line given by  (\ref{eq:smallupot}).}
\label{fig:EffPot}
\end{center}
\end{figure}

\paragraph{Conformal sector.} 

The existence of a critical value for $u$ forces us to search for a new vacuum solution other than those given by \eqref{eq:lambdavaccmassless}. Indeed, \figref{fig:PotentialsM0} clearly shows a new vacuum located at $\sigma=0$ which survives at nonzero heterotic deformations. For arbitrary value of $u$ equations \eqref{eq:VacuaEqnsxs} admit the following solution 
\[
|\sigma_{0}|=0, \quad D=0\,.
\label{eq:zerovaccmassless}
\]
This solution formally exists for the $\CP{N-1}$ sigma-model as well, but in that case it must be discarded. As can be seen in \figref{fig:PotentialsM0}, it represents a maximum, rather than a minimum. Strictly speaking, the effective potential cannot be trusted for $\sigma=0$, where some degrees of freedom become massless.  {The existence of massless kinks is ensured once we interpret the vacuum at $\sigma=0$ as a degenerate point where $\Nt$ vacua coalesce.  We will check this explicitly in the next section, where we will resolve the $\Nt$ vacua by the introduction of twisted masses  \cite{Dorey:1999zk,Dorey:1998yh,Witten:1993yc}. This sector of the theory is described by a super-conformal field theory. This was first conjectured in Refs. \cite{Dorey:1999zk,Shifman:2006bs}, by analogy with the four dimensional case: coalescence of vacua in two-dimensional theories corresponds to the degeneration of Seiberg-Witten curves \cite{Seiberg:1994rs,Seiberg:1994aj} at the so-called Argyres-Douglas points of four-dimensional theories\footnote{The four-dimensional curve $y^{2}=f(x,\Lambda)$ is given in terms of the two-dimensional superpotential: $y^{2}=(\partial W(x)/\partial x)^{2}$. Using \eqref{eq:masslesssuperpot} we have $y^{2}=x^{2\Nt}(x^{N-\Nt}-\Lambda^{N-\Nt})^{2}$, which has $\Nt$ degenerate singularities.}, where the appearance of massless, mutually non-local degrees of freedom gives rise to an interacting super-conformal field theory \cite{Argyres:1995jj,Argyres:1995xn}. This expectation was confirmed in Ref. \cite{Tong:2006pa}, where it was shown that the two-dimensional theory flows to an interacting  super-conformal fixed point, identified as an $A_{N-1}$ minimal model \cite{Martinec:1988zu,Vafa:1988uu,Witten:1993jg}, as $\sigma \rightarrow 0$\footnote{Ref. \cite{Tong:2006pa} deals with the case of complete degeneration of the vacua: $y^{2}=x^{2N}$. The qualitative aspects of that analysis hold in our case as well.}.  Notice that we can trust both the large-$N$ effective potential and the exact twisted superpotential for arbitrarily small $\sigma$ as soon as we interpret them as valid at energy scales $\epsilon$ much smaller than the masses of the hypermultiplets $\epsilon\ll |m_{hyp}|\sim |\sigma|$ (see \eqref{eq:complagr}). The divergences of both potentials arise because of infrared instabilities due to the developing of massless states, as described above.


If we assume continuity of physical quantities in the limit $\sigma \rightarrow 0$, the result of this section (see \figref{fig:PotentialsM0}) implies that this super-conformal sector is not lifted by the heterotic deformation. Furthermore, supersymmetry is not broken for $\sigma=0$. The massive vacua discussed in this section become  metastable when we turn on $u$, and disappear as we increase the heterotic deformation above the critical value $u_{crit}$.} 

Let us conclude this section by analytically solving \eqref{eq:VacuaEqnsxs}  for small values of $u$. As we see from \figref{fig:UDomainM0} this also implies small $x$.  One thus has from \eqref{eq:VacuaEqnsxs}
\<\label{eq:VacuaEqnsxsexp}
(1+x)(1-\alpha x) \eq 1+u\,,\nln
\left(1-{\frac{\alpha}{1-\alpha }}x\right) \left(1+{\frac{1}{\alpha-1}}x\right) \eq s\,,
\>
which gives
\[
x \approx \frac{u}{1-\alpha}\,, \quad s \approx 1-\frac{1+\alpha}{(1-\alpha)^2}u\, .
\label{eq:smallsol}
\]
Substituting \eqref{eq:smallsol} in the expression for the effective potential \eqref{eq:EffectivePotentialMassless } we find the  the following expression for the vacuum energy
\<
V_{eff}\eq \frac{N}{4\pi}u\Lambda^2 \,,
\label{eq:smallupot}
\>
which is to be compared with the numerical solution in \figref{fig:EffPot}. Notice that the small $u$ limit does not depend on the value of $\alpha$.

\subsection{Massive case}

We shall now determine the vacuum structure of the massive model in terms of the two-dimensional space of parameters $m$ and $\mu$.  With a quick inspection to the first line of \eqref{eq:VacEqDsigma}
\<
\left(\left|\sigma- m_0\right|^2+D\right)n_0=0\,,\quad \left(\left|\sigma- \mu_0\right|^2-D\right)\rho_0=0\,, 
\nonumber \\
\>
we can easily identify three branches of solutions,  which correspond to three different phases of the theory
\begin{enumerate}
\item[$\textbf{H}{n}$]: Higgs phase with non-zero VEV for  $n^i$
\[\label{eq:PhaseHC}
\rho_0 =0\,, \quad D=-\left|\sigma- m_0\right|^2\,,
\]
\item[$\textbf{H}{\rho}$]: Higgs phase with non-zero VEV for  $\rho^j$
\[\label{eq:PhaseCH}
n_0=0\,\quad D=\left|\sigma- \mu_0\right|^2\,,
\]
\item[\textbf{C}]: Coulomb phase
\[\label{eq:PhaseCC}
n_0 = \rho_0 =0\,.
\]
\end{enumerate}
Recall that the renormalized coupling is given by \eqref{eq:rencoupling}, $r=|n_{0}|^{2}-|\rho_{0}|^{2}$. Thus, the $\textbf{H}{n}$ phase is characterized by a positive coupling, while  in the $\textbf{H}{\rho}$ phase the renormalized  Fayet-Iliopoulos term is negative. In the \textbf{C} phase $r=0$. We will determine the appearance of these phases in the $m-\mu$ plane, by starting with the undeformed case.

\subsubsection{Undeformed case}

The $\ssN=(2,2)$ sigma-model is solved by virtue of the exact superpotential. As we have mentioned in the introduction, in the current paper we shall work with the large-$N$ approximation as it can be used both for the $(2,2)$ and $(0,2)$ models.

\begin{figure}
\begin{center}
\includegraphics[height=12cm, width=12cm]{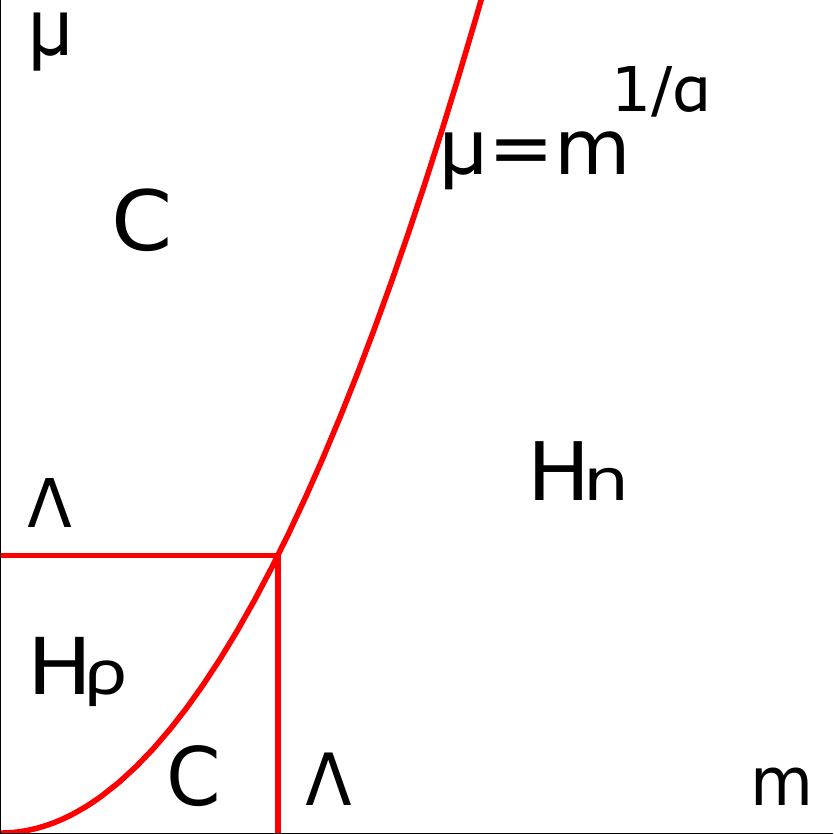}
\caption{Phase Diagram of the weighted $(2,2)$ $\mathbb{CP}^{N-1}$ model in the large-$N$ approach. There are four domains with different VEVs for $\sigma$: two Higgs branches $\textbf{H}{\rho}$ and  $\textbf{H}{n}$, and two Coulomb branches \textbf{C}. In the Coulomb phase \textbf{C}  $r=0$. The curve $\mu/\Lambda=(m/\Lambda)^{1/\alpha}$ together with horizontal and vertical lines starting from $\mu=\Lambda$ and $m=\Lambda$ respectively separates the \textbf{C} phases from the Higgs phases. In $\textbf{H}{n}$ $r>0$ and in $\textbf{H}{\rho}$ $r<0$. On the super-conformal line $\mu/\Lambda=(m/\Lambda)^{1/\alpha}$ a new branch described by a super-conformal theory opens up.}
\label{fig:PhaseDiagram22}
\end{center}
\end{figure}

We anticipate here the discussion of this section by proposing the phase diagram in \figref{fig:PhaseDiagram22}. Below we list the vacua solutions in each domain of the phase diagram.
%

\paragraph{$\textbf{H}{n}$ phase.}
The unbroken supersymmetry of the undeformed model implies $D=0$ for all the phases. From \eqref{eq:PhaseHC} we thus find
\[
\sigma = m_0\,,\quad \rho_{0}=0, \quad D=0\,.
\]
From the second line of  \eqref{eq:VacEqDsigma} we determine the coupling constant
\[\label{eq:CouplingRen}
r = |n_0|^2 = \frac{1}{4\pi}\sum\limits_{i=1}^{N-1} \log\frac{\left|m_0-m_i\right|^2}{\Lambda^2} -
\frac{1}{4\pi}\sum\limits_{j=1}^{\tilde N-1} \log\frac{\left|m_0-\mu_j\right|^2}{\Lambda^2}\geq 0\,.
\]
The sums in the expression above can be exactly calculated in the large-$N$ limit as shown in Ref. \cite{Bolokhov:2009wv}
\<\label{eq:CouplingRenHC}
r =
\left\{
\begin{array}{ll}
  \frac{N-\tilde N}{2\pi}\log\frac{m}{\Lambda}\,, & \mu<m     \\
  & \\
  \frac{N}{2\pi}\log\frac{m}{\Lambda}-\frac{\tilde N}{2\pi}\log\frac{\mu}{\Lambda}\,, & \mu>m\,.
\end{array}
\right. \nonumber \\
\>
By asking for $r$ to be positive, we obtain  the following conditions for the existence of the $\textbf{H}{n}$ phase
\<
\label{eq:HClineineq}
\textbf{H}{n}: \quad\left\{
\begin{array}{ll}
  m  >  \Lambda , & \mu<m     \\
  & \\
 \frac{m}{\Lambda} >   \left(\frac{\mu}{\Lambda}\right)^{\alpha} ,  & \mu>m\,.
\end{array}
\right. \nonumber \\
\>
%

\paragraph{$\textbf{H}{\rho}$ phase.}

In this phase we use \eqref{eq:PhaseCH} to find
\[
\sigma = \mu_0\,,\quad n_{0}=0, \quad D=0\,,
\]
and the coupling constant
\<\label{eq:CouplingRenCH}
r =
\left\{
\begin{array}{ll}
  \frac{N-\tilde N}{2\pi}\log\frac{\mu}{\Lambda}\,, & \mu>m     \\
  & \\
  \frac{N}{2\pi}\log\frac{m}{\Lambda}-\frac{\tilde N}{2\pi}\log\frac{\mu}{\Lambda}\,, & \mu<m
\end{array}
\right. \nonumber \\
\>
Negativity of $r$ now implies the following conditions for the existence of the $\textbf{H}{\rho}$ phase
\[\label{eq:CHlineineq}
 \textbf{H}{\rho}: \quad    \left(\frac{m}{\Lambda}\right)^{1/\alpha} < \frac{\mu}{\Lambda}< 1\,.
\]

The renormalized coupling constant vanishes, as expected, along the boundaries of the Higgs phases. As we will explain later the curve defined as
\[\label{eq:HCCHline}
\frac{m}{\Lambda} = \left(\frac{\mu}{\Lambda}\right)^\alpha\,,
\]
is of particular interest.
Notice that the renormalized coupling in both Higgs regimes scales with $N$. Let us mention that a more natural coupling constant would be
\[\label{eq:tHooftlike}
\ell=\frac{1}{\lambda} = \frac{4\pi r}{N},
\]
which is reminiscent of the 't-Hooft coupling constant which naturally appears in large-$N$ gauge theories.

\paragraph{\textbf{C} phases.}

The Coulomb phase exists in the regions where  $|n_{0}|=|\rho_{0}|=0$. There are two distinct regions, \textbf{C}$\mu$ and  \textbf{C}$m$, which complete the phase diagram as shown in \figref{fig:PhaseDiagram22}. With the first and the third equations of \eqref{eq:VacEqDsigma} being satisfied automatically for $u=D=0$, the second one gives
\[
\prod\limits_{i=1}^{N-1}|\sigma - m_i|^2 = \Lambda^{N-\tilde N}\prod\limits_{j=1}^{\tilde N-1}|\sigma - \mu_j|^2\,.
\]
Note that each  part of this equation is real for the complex variable $\sigma$. This implies a continuous set of solutions which are located on a closed line. Again, this is the effect of the large-$N$ approximation, where an infinite number of vacua is continuously distributed on a curve. The solution in the leading approximation is qualitatively different in the two \textbf{C} regions. In the \textbf{C}$\mu$ region the vacua sit on a single circle
\<
\label{eq:CCmuvacua}
|\sigma_{\mu}|= \Lambda \left(\frac{\mu}{\Lambda}\right)^{\alpha}\,,\quad  \quad \quad \mu>\Lambda \left(\frac{m}{\Lambda}\right)^{1/\alpha}, \quad \mu>\Lambda\,.
 \>
In the \textbf{C}$m$ region the vacua split between two separate circles
\[\label{eq:CCVacua}
\begin{array}{l}
 |\sigma_{m}| = \Lambda \left(\frac{m}{\Lambda}\right)^{1/\alpha}\,,   \\
 \\
|\sigma_{\Lambda}| = \Lambda\,
\end{array}  
\quad \quad\mu<\Lambda \left(\frac{m}{\Lambda}\right)^{1/\alpha},\, \quad m<\Lambda\,.
\]
In order to resolve the vacua into a discrete set we can  compare the result above with the exact one  given by the $\ssN=(2,2)$ superpotential. Vacua are the solutions of the equation \cite{Dorey:1999zk}
\[
\prod\limits_{i=1}^{N-1}(\sigma - m_i)=\Lambda^{N-\tilde N}\prod\limits_{j=1}^{\tilde N-1}(\sigma - \mu_j)\,,
\]
which, making use of Eq. \eqref{eq:CircleMasses}, can be rewritten as the following: 
\[
\sigma^N -m^N= \Lambda^{N-\tilde N}(\sigma^{\tilde N} -\mu^{\tilde N})\,.
\label{eq:exactvacua}
\]
It is exact even for small $N$, but in the large-$N$ approximation one obtains three groups of solutions: $N$ ``$\mu$-vacua'' in the \textbf{C}$\mu$ region
\<
\sigma_{\mu,j}=\Lambda \left(\frac{\mu}{\Lambda}\right)^{\alpha}e^{2\pi i \frac{j}{ N}} \quad \quad j=0,\dots,N-1\,,  \nonumber \\
\>
while in the \textbf{C}$m$ we have  $N-\tilde N$ ``$\Lambda$-vacua'' 
\[
\sigma_{\Lambda,k} = \Lambda\, e^{2\pi i \frac{k}{N-\tilde N}}\,,\quad k = 0,\dots,N-\tilde N-1\,,
\]
and $\tilde N$ ``$m$-vacua''
\<
\label{eq:CoulombSmallVacua}
\sigma_{m,l} =
  \Lambda \left(\frac{m}{\Lambda}\right)^{1/\alpha}e^{2\pi i \frac{l}{\tilde N}}\,,   \quad \quad l=1,\dots,\Nt-1\,.
\>
%

\paragraph{Super-conformal line.}

{  The following special situation occurs on the line 
\[
\frac{\mu}{\Lambda}=\left(\frac{m}{\Lambda}\right)^{1/\alpha}\,,
\label{eq:LineSCFT}
\]
where \eqref{eq:exactvacua} degenerates to 
 \[
\sigma^N = \Lambda^{N-\tilde N}\sigma^{\tilde N}\,.
\label{eq:exactvacuared}
\]
This equation has two sets of solutions
\[
\sigma^{N-\tilde N} = \Lambda^{N-\tilde N}, \quad \sigma=0\,,
\]
where the latter solution applies to the conformal regime. Recall that we had a similar situation in the massless case. For this particular configuration we obtain $N-\Nt$ massive vacua and a sector where  $\sigma$ vanishes. The same considerations made for the massless case apply in the massive case on the whole super-conformal line\footnote{Notice also that the Seiberg-Witten curve of the corresponding four-dimensional theory 
$
y^{2}=x^{N}-m^{N}-\Lambda^{N-\Nt}(x^{\Nt}-\mu^{\Nt})^{2}\,,
$
provided that \eqref{eq:LineSCFT} holds, is reduced to 
$
y^2 = x^{2\Nt}(x^{N-\Nt}-\Lambda^{N-\Nt})^{2}\,
$ \cite{Tong:2006pa},
i.e. it has the same form along the whole super-conformal line.
}}.

\subsubsection{Small heterotic deformation}

We shall now introduce the heterotic deformation in the model.
Let us first study corrections to the vacuum expectation values of our fields for small $u$. 

\paragraph{$\textbf{H}{n}$ phase.}

We can easily solve the first and second equations of \eqref{eq:VacEqDsigma} for  $D$ and $|n_{0}|^{2}$. The third line is thus an equation for $\sigma$
\begin{align}
\label{eq:EqSigmaHiggs}
 &\sum\limits_{i=1}^{N-1}\left(\sigma- m_i\right)\log\left(1-\frac{|\sigma- m_0|^2}{|\sigma- m_i|^2}\right)+   \sum\limits_{j=1}^{\tilde N-1}\left(\sigma- \mu_j\right)\log\left(1+\frac{|\sigma- m_0|^2}{|\sigma- \mu_j|^2}\right) +  \notag \\
 & - \left(\sigma- m_0\right) \,  r =  u N\sigma\,.
\end{align}
We shall now expand the equation above in terms of small deviations from the undeformed case
\<
\sigma=m_{0}+\delta\sigma, \quad D=0+\delta D, \quad r=r_{0}+\delta r\,,
\>
which is a consistent procedure when $u$ is small.  Equation  \eqref{eq:EqSigmaHiggs}  gives the following result
\[
\delta \sigma=-m\frac{uN}{ r_{0}}\,,
\]
where $r_0$ is given by \eqref{eq:CouplingRenHC}. The correction to $D$ can be easily found to be
\[
\delta D=-|\delta\sigma|^{2}\,.
\]
Finally we write the expression for the correction to the renormalized coupling constant
\<\label{eq:rCorrectionHiggs}
\delta r \eq  \frac{N}{2\pi}\delta \sigma\left(\frac{1}{N}\sum\limits_{i=1}^{N-1}\frac{2 \Re(m_{0}-m_{i})}{|m_0-m_i|^{2}}-\alpha\frac{1}{\tilde N}\sum\limits_{j=1}^{\tilde N-1}\frac{2 \Re(m_{0}-\mu_{j})}{|m_0-\mu_j|^{2}}\right)\nln
\eq -\frac{N u}{2 \pi r_{0}}\left(1-\alpha f\left(\frac{\mu}{m}\right)\right)\,,
\>
where 
\[\label{eq:fofbeta}
f(\beta)=
\left\{ 
\begin{array}{cc} 
2\,,&\quad \beta<1 \\ 
1 \,,&\quad \beta=1 \\ 
0\,,&\quad \beta>1\,. 
\end{array} 
\right.
\]
The last equality holds in the large-$N$ limit. Notice that all the corrections contain a $1/r_{0}$ factor. They all diverge as we approach the Coulomb phase boundary, when $r\rightarrow 0$. In this region our approximation fails. Nonetheless, in the  boundary region with the Coulomb phase we have $\mu > m$ and  the correction is negative, thereby reducing the value of $r$. We can argue that the \textbf{H}$n$ phase gets shrunk. This expectation will be confirmed further in the study of the large $u$ case.
%
%

\paragraph{$\textbf{H}{\rho}$ phase.}

We can proceed analogously to the $\textbf{H}{n}$ phase, obtaining
\[
\delta \sigma =\mu\frac{uN}{r_{0}}\,,
\]
whereas the correction to the coupling reads
\[\label{eq:rCorrectionCH}
\delta r = \frac{ N u}{2 \pi r_{0}}\left(f\left(\frac{m}{\mu}\right)-\alpha\right)\,.
\]
The same comments holds for this phase. In particular, the correction near the boundary with the Coulomb phase $\mu<m$ is positive, thus it is plausible that the second \textbf{H}$\rho$ region is also reduced.

\paragraph{\textbf{C} phase.}

In this phase both $n_0$ and $\rho_0$ vanish. We only need \eqref{eq:VacEqDsigma} to determine the correction to the VEV of $\sigma$. The second equation of \eqref{eq:VacEqDsigma} now is
\[\label{eq:DVarCoulomb}
\prod\limits_{i=1}^{N-1}(|\sigma- m_i|^2+D)=\Lambda^{N-\tilde N}\prod\limits_{j=1}^{\tilde N-1}(|\sigma- \mu_j|^2-D)\,,
\]
while the third one gives
\<\label{eq:SigmaVarCoulumb}
\sum\limits_{i=1}^{N-1} \left(\sigma- m_i\right)\log\left(1+\frac{D}{\left|\sigma- m_i\right|^2}\right) \nln
+\sum\limits_{j=1}^{\tilde N-1} \left(\sigma- \mu_j\right)\log\left(1-\frac{D}{\left|\sigma- \mu_j\right|^2}\right) \eq  N u \sigma\,.
\>
We look again for the solution of the form $\sigma = \sigma_{0}+\delta\sigma$. 
From \eqref{eq:SigmaVarCoulumb} we get
\<
\label{eq:DCpulomb}
\delta D= u|\sigma_{0}|^{2} \, , &\quad& m<|\sigma_{0}|<\mu\,,\nln
\delta D=-\frac{u}{\alpha}|\sigma_{0}|^{2}\, , &\quad& \mu<|\sigma_{0}|<m\,,\nln
\delta D=\frac{u}{1-\alpha}|\sigma_{0}|^{2}\, , &\quad& \mu,m<|\sigma_{0}|\,.
\label{eq:smallDvalues}
\>
From \eqref{eq:DVarCoulomb} we can find the correction to $\sigma_{0}$. Expanding this equation we get
\<
N(\sigma_{0}^{N}-m^{N})^{2}(\delta\sigma\sigma_{0} f(m/\sigma_{0})+\delta D g(m/\sigma_{0}))= & & \nonumber \\
 =  \Nt\Lambda^{N-\Nt} (\sigma_{0}^{\Nt}-\mu^{\Nt})^{2}(\delta\sigma \sigma_{0}f(\mu/\sigma_{0})-\delta D g(\mu/\sigma_{0})),& & \nonumber \\
\label{eq:deltasigmaexp}
\>
where $f(\beta)$ is defined in \eqref{eq:fofbeta} and $g(\beta)$ is \cite{Bolokhov:2009wv}
\[
g(\beta) =\frac{1}{N}\sum\limits_{k=1}^{N}\frac{1}{|1-\beta e^{2 \pi i k/N}|^{2}}=  \frac{1}{|1-\beta^{2}|}\,.
\]

We are now ready to write down the results for the two Coulomb regions. First, in the \textbf{C}$\mu$ region, from \eqref{eq:CCmuvacua} we have $\mu>\sigma_{\mu,0}>m$; information that we need to correctly evaluate \eqref{eq:deltasigmaexp}. Using also the first line of \eqref{eq:smallDvalues} we obtain for the  $\mu$-vacua
\<
\label{eq:CoulombSmallVacuaCorrectedmu}
\textbf{C}\mu: \quad |\sigma_{\mu}| =
 \Lambda \left(\frac{\mu}{\Lambda}\right)^{\alpha}\left(1- \frac{u/2}{1-m^{2}/\sigma_{\mu,0}^{2}}+\alpha \frac{u/2}{1-\mu^{2}/\sigma_{\mu,0}^{2}}   \right)\,.
 \>
In the \textbf{C}$m$ we need to find the corrections for both the $m$-vacua and the $\Lambda$-vacua. Using the right values of $\sigma$  and the second and third line of  \eqref{eq:smallDvalues} we get respectively
\<
\label{eq:CoulombSmallVacuaCorrected}
\textbf{C}m: \quad \begin{array}{l}
|\sigma_{m}|=   \Lambda \left(\frac{m}{\Lambda}\right)^{1/\alpha}\left(1-\frac{u}{2\alpha}\frac{1}{1-\mu^{2}/\sigma_{m,0}^{2}}+\frac{u}{2\alpha^{2}} \frac{1}{1-m^{2}/\sigma_{m,0}^{2}}   \right)\,,  \\
   \\
 |\sigma_{\Lambda}|=\Lambda \left(1-\frac{u}{2(1-\alpha)^{2}}\frac{1}{1-m^{2}/\Lambda^{2}}-\frac{u\alpha}{2(1-\alpha)^{2}}\frac{1}{1-\mu^{2}/\Lambda^{2}}   \right)\,.
\end{array}
\>

Let us make a couple of comments. The presented calculation is valid in the large-$N$ limit for all values of masses $m$ and $\mu$ in the two Coulomb regions. All the corrections are negative, thus they reduce the VEV of $\sigma$. The calculation breaks down at the boundary of the Coulomb regions with the Higgs phases. Notice that the value of $D$ tends to zero if we approach the massless case (we can reach it, for example, through the \textbf{C}$m$  phase along the $\mu=0$ axis). This is a strong hint that the theory in the super-conformal regime does not break supersymmetry. Unfortunately, it is not possible to use this small-$u$ expansion to check the same result for the whole super-conformal line, where the corrections calculated above diverge\footnote{We will be able to prove this in the large $u$ limit in the next section.}.  The factor $1-\alpha$ in the expression for the $\Lambda$-vacua arises naturally if we recall that the theory is conformally invariant for $\alpha=1$. In this limiting case there are no $\Lambda$-vacua. Moreover, the result is consistent with the expectation that a critical value of $u$ appears in the massive case such that the $\Lambda$ vacua disappear at larger values of $u$. As in the massless case, the value of $u_{crit}$ should tend to zero as $\alpha$ approaches $1$. 

For small $u$, the vacuum energy is simply given by the heterotic deformation in the potential: 
\[
E=\frac{uN}{4 \pi}|\sigma_{0}|^{2}\,.
\label{eq:smallenergy}
\]
As expected, vacuum energy is thus larger for larger values of the VEV of $\sigma$. While in the \textbf{C}$\mu$ region all the vacua have the same energy, in the \textbf{C}$m$ phase the $\Lambda$-vacua acquire a much larger energy, as compared to the $\mu$-vacua. As was noticed in the massless case, $\Lambda$-vacua become metastable once we turn on the heterotic deformation. In the next section we consider the large $u$ limit and we will assume that $\sigma$ is always small in the vacuum, which is a consistent assumption once the  $\Lambda$-vacua have ceased to exist for a sufficiently large $u$.

\subsubsection{Large heterotic deformation}

At generic values of the deformation parameter $u$ we can only rely on numerical solutions of the full equations \eqref{eq:VacEqDsigma}. Unfortunately, this is quite complicated. In this section we will simplify the problem by looking at large values of $u$. Then we will compare the results with some full numerical calculations done at generic values of $u$, as a double check of both results.
\begin{figure}
\begin{center}
\includegraphics[height=15cm, width=14.5cm]{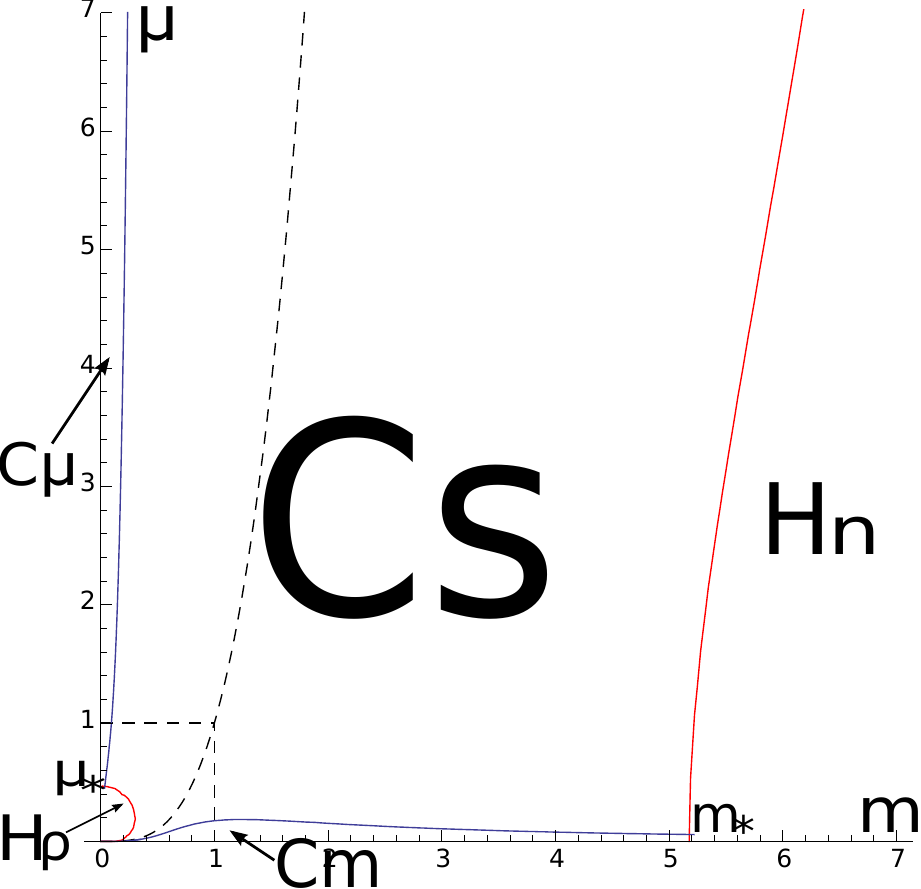}
\caption{Phase Diagram for  $u=10$ and $\alpha=.3$.  At $u=0$ the \textbf{C} phase of the theory had non-zero VEV for $\sigma$ everywhere but on the graph $\mu/\Lambda = (m/\Lambda)^{1/\alpha}$. As we increase $u$ the domain with unbroken $\Integers_{\tilde N}$ symmetry (\textbf{Cs} phase) gets widened pushing \textbf{C} phases with broken symmetry towards the axes. At very large values of $u$ the latter phase occupies only two small domains as is shown in the figure. The Higgs phases $\textbf{H}{n}$ and $\textbf{H}{\rho}$ are also pushed apart by blowing \textbf{Cs} phase, one can see it from (\ref{eq:mast}) and (\ref{eq:muast}). In the limit $u\to \infty$ the theory has only \textbf{Cs} phase. To show all the phases, we magnified \textbf{C}$m$ region by a factor of $10^{7}$ and the \textbf{C$\mu$} by a factor of 20. }
\label{fig:PhaseDiagLargeu}
\end{center}
\end{figure}
%

\paragraph{$\textbf{H}{n}$ phase.}

As noted in Ref. \cite{Bolokhov:2010hv} the large $u$  approximation can be exploited by considering $\sigma\ll m$. Finding $D$ from the first line of \eqref{eq:VacEqDsigma} and substituting it in the third line, if we ignore $\sigma$ compared to $m$ , we get (we also ignore terms enhanced by the logarithms)
\[
\frac{4\pi}{N} r-\log \left(\frac{4\pi}{N} r+1\right)=\log\frac{m^2}{u\Lambda^2}-\alpha\log\frac{m^2+\mu ^2}{\Lambda^2}\,,
\]
where $r$ is given by
\[
r=\frac{N}{4\pi}\log\left( \frac{\sigma m}{\Lambda^{2}}\right)-\frac{\Nt}{4 \pi}\log\left(\frac{m^{2}+\mu^{2}}{\Lambda^{2}}\right)\,.
\]
We can now find the boundary of this branch with the Coulomb branch of the theory by forcing $r=0$ in the above equation. It gives us
\[\label{eq:HCboundary}
\left(\frac{m}{\Lambda}\right)^2 = u\left(\frac{m^2+\mu ^2}{\Lambda^2}\right)^\alpha\,.
\]
This boundary is shown in \figref{fig:PhaseDiagLargeu}. One can see that it gets shifted towards the large values of $m$ as $u$ increases. The value of the phase transition point on the  $\mu=0$ axis is
\[
m_\ast = \Lambda u^{\frac{1}{2-2\alpha}}\,.
\label{eq:mast}
\]
%

\paragraph{$\textbf{H}{\rho}$ phase.}
The procedure in this phase is similar. Now we exploit the approximation $\sigma\ll\mu$ valid in the large $u$ limit. The equation for the renormalized coupling is
\[
\frac{4\pi}{N} r+\alpha \log \left(\frac{4\pi}{N} r+\alpha\right)=\log \frac{m^2+\mu^2}{\Lambda^2}-\alpha\log\frac{\mu ^2}{u\Lambda^2}\,,
\]
where $r$ is given by
\[
r=-\frac{\Nt}{4\pi}\log\left( \frac{\sigma \mu}{\Lambda^{2}}\right)+\frac{N}{4 \pi}\log\left(\frac{m^{2}+\mu^{2}}{\Lambda^{2}}   \right)
\]
The boundary between the  $\textbf{H}{\rho}$ and Coulomb is parametrized by the following equation
\[\label{eq:CHboundary}
\left(\frac{\mu}{\Lambda}\right)^2 = \frac{u}{\alpha}\left(\frac{m^2+\mu ^2}{\Lambda^2}\right)^{1/\alpha}\,.
\]
This boundary is shown in \figref{fig:PhaseDiagLargeu}. The phase transition for $m=0$ occurs at
\[
\mu_\ast = \Lambda\left(\frac{\alpha}{u}\right)^{\frac{\alpha}{2-2\alpha}}\,.
\label{eq:muast}
\]
\begin{figure}[ht]
\includegraphics[height=5cm, width=7.7cm]{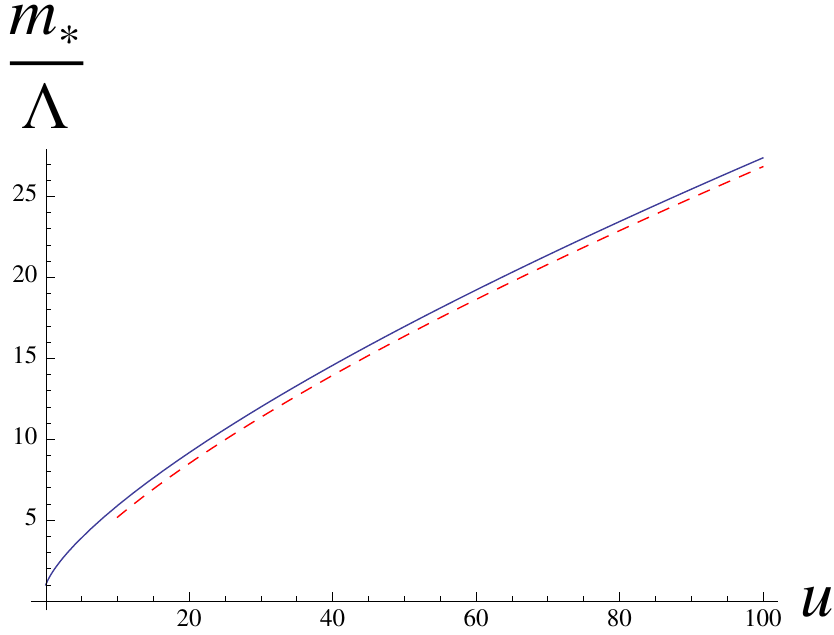}  \quad
\includegraphics[height=5cm, width=7.7cm]{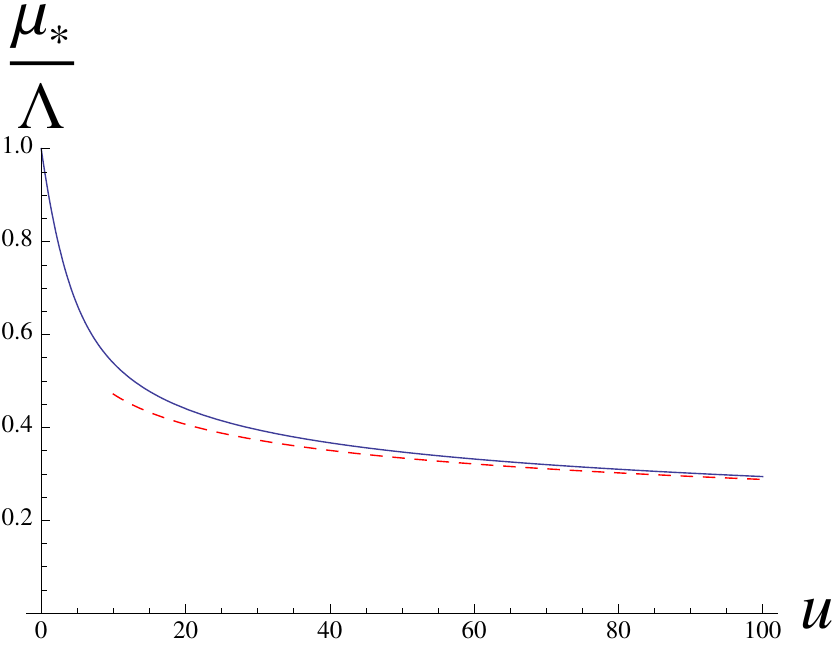}\\
\caption{Dependence of $m_\ast$ (left plot) and $\mu_\ast$ (right plot) as a function of $u$. Here we compare numerical solutions (solid lines) and the analytical values (\ref{eq:mast}) and (\ref{eq:muast}) (dashed lines) for  $\alpha=0.3$.}
\label{fig:massstar}
\end{figure}

\paragraph{\textbf{C} phases.}
%
%
%

In Ref. \cite{Bolokhov:2010hv} a new phase within the Coulomb regime at large $u$ was found where  $\corr{\sigma}=0$ and the residual discrete symmetry was not broken. For the masses which are exponentially small in $u$, a VEV for $\sigma$ is restored, and a Coulomb phase with broken symmetry appears.
To study how their picture is generalized for the weighted sigma-model, we search for a broken Coulomb phase in the two following regions
\<
 &\textbf{C}\mu\,:& m\,\ll\,\mu, \nonumber \\
&\textbf{C}m\,:& \mu\,\ll\,m,
\>
where we shall use the following assumptions\footnote{These assumptions are justified because we search for exponentially small values of $\sigma$ and $m$ (or $\mu$) and we expect $D$ to be large for large $u$.}
\<
 &\textbf{C}\mu\,:& \sigma,\,m\,\ll\,\mu,\,D \nonumber \\
&\textbf{C}m\,:& \sigma,\,\mu\,\ll\,m,\,D\,.
\label{eq:CCAssumpt}
\>

Let us start with the \textbf{C}$\mu$ phase. By employing the approximations \eqref{eq:CCAssumpt} in \eqref{eq:VacEqDsigma} we get the following equations
\<
D  =   \Lambda^2\left(\left(\frac{\mu}{\Lambda}\right)^2-\frac{D}{\Lambda^2}\right)^{\alpha}\,, \nonumber &&  \\
2 \alpha \sigma^{2}\log\left(\frac{\mu^{2}-D}{\Lambda \mu} \right)-u \sigma^{2}&=&
\left\{
\begin{array}{cc}
 2 \sigma^{2}\log\left(\frac{\sigma}{\Lambda} \right)  &  m<\sigma  \\
  2 \sigma^{2}\log\left(\frac{m}{\Lambda} \right) &  m>\sigma  
\end{array}
\right.
\label{eq:CCmuappr}\,.
\>

Notice that the equations above admit only the solution $\sigma=0$ as long as $m>\sigma$. This is the Coulomb symmetric phase. For smaller $m$, we pick the first line in \eqref{eq:CCmuappr}, wich gives non-trivial values for $\sigma$. We can actually determine the boundary \textbf{C}$\mu$-\textbf{C}s by solving the above equations for $m<\sigma$ and then imposing the condition $\sigma=m$.  The \textbf{C}$\mu$ phase gets extended towards smaller values of $\mu$, when it will eventually meet the $\textbf{H}{\rho}$ phase. As a final check, let us further simplify \eqref{eq:CCmuappr} in the large $\mu$ limit
\<
D  \eq    \Lambda^2\left(\frac{\mu}{\Lambda}\right)^{2\alpha}\,, \nln
\sigma \eq \Lambda \left(\frac{\mu}{\Lambda}\right)^{\alpha}e^{-u/2}\,,
\label{eq:CCmuapprlargemu}
\>
which is consistent with the results of Ref. \cite{Bolokhov:2010hv}. In this region, in fact, our model reduces to the ordinary $\CP{N-1}$ model, with the new scale $\tilde \Lambda=\Lambda (\mu/\Lambda)^{\alpha}$.

The \textbf{C}$m$ phase is completely analogous. The correct approximation leads us now to the following equations
\<
m^{2}+D  =   \Lambda^2\left(-\frac{D}{\Lambda^2}\right)^{\alpha}\,, \nonumber   \\
2  \sigma^{2}\log\left(\frac{m^{2}-D}{\Lambda m} \right)-u \sigma^{2}&=&
\alpha \left\{
\begin{array}{cc}
 2 \sigma^{2}\log\left(\frac{\sigma}{\Lambda} \right)  &  \mu<\sigma  \\
  2 \sigma^{2}\log\left(\frac{\mu}{\Lambda} \right)  &  \mu>\sigma  
\end{array}
\right.
\label{eq:CCmappr}\,.
\>
We proceed as for the \textbf{C}$\mu$ phase to determine the boundary with the symmetric phase. The result is shown in \figref{fig:PhaseDiagLargeu}. If we look at very small values of $m$, we can simplify \eqref{eq:CCmappr} a bit more. From the first equation we see that very small $m$ implies $m^{2}\gg D$. Finally we get
\<\label{eq:CCmapprsmallm}
-D \eq    \Lambda^2\left(\frac{m}{\Lambda}\right)^{2/\alpha}\,, \nln
\sigma \eq \Lambda \left(\frac{m}{\Lambda}\right)^{1/\alpha}e^{-u/(2\alpha)}\,.
\>

\paragraph{Super-conformal line.}

Keeping the results of this section in mind, it is  now easy to check that supersymmetry is effectively unbroken as we approach the  super-conformal line 
\[\label{eq:SuperConfLine2}
\frac{\mu}{\Lambda}=\left(\frac{m}{\Lambda}\right)^{1/\alpha}\,.
\]
Since we are looking into the \textbf{C}$s$ phase, we put from the beginning $r=0$ and $\sigma=0$ in the second line of\eqref{eq:VacEqDsigma}
\[
\left(m^{2}+D\right)^N=\Lambda^{N-\tilde N}\left(\mu^{2}-D\right)^{\tilde N}\,,
\]
which is clearly solved by $D=0$ provided that \eqref{eq:SuperConfLine2} holds. This condition is enough to show unbroken supersymmetry. One can also directly check that the vacuum energy vanishes. In general, in the \textbf{C}$s$ phase $D$ does not vanish,  and supersymmetry is generically broken.

\section{Spectrum}\label{Sec:Spectrum}

As was shown by Witten in the supersymmetric $\CP{N-1}$ sigma-model photon is massive due to a coupling to fermions and its mass is given by the chiral anomaly \cite{Witten:1978bc}. However, the photon remains massless in the bosonic $\CP{N-1}$ sigma model. It was shown in Ref. \cite{Bolokhov:2010hv} that once the twisted masses are nonzero and the heterotic deformation is turned on, the photon becomes massless in the symmetric Coulomb phase. The authors also call this phase confining, since existence of long range interactions with massless carrier allows bound states of particles (``kinks''). In $\CP{N-1}$ sigma-model only $\bar{n}n$ mesons could be formed, our model also admits, in principle, $\bar{\rho}\rho$ and $n\rho$ mesons. Below we calculate the photon mass at different values of twisted masses $m$ and $\mu$ as well as the heterotic deformation parameter $u$, and show that it vanishes in the symmetric Coulomb phase as is prescribed by the unbroken discrete symmetry. Since analogous calculations in supersymmetric sigma-models have been previously performed (see, for instance Refs. \cite{Shifman:2008kj, Bolokhov:2010hv, Koroteev:2010gt}) here we shall just list our result. Generic expressions for the effective coupling constants can be found in \secref{Sec:SomeFormulae}.

\begin{figure}
\begin{center}
\includegraphics[height=2.4cm, width=10.5cm]{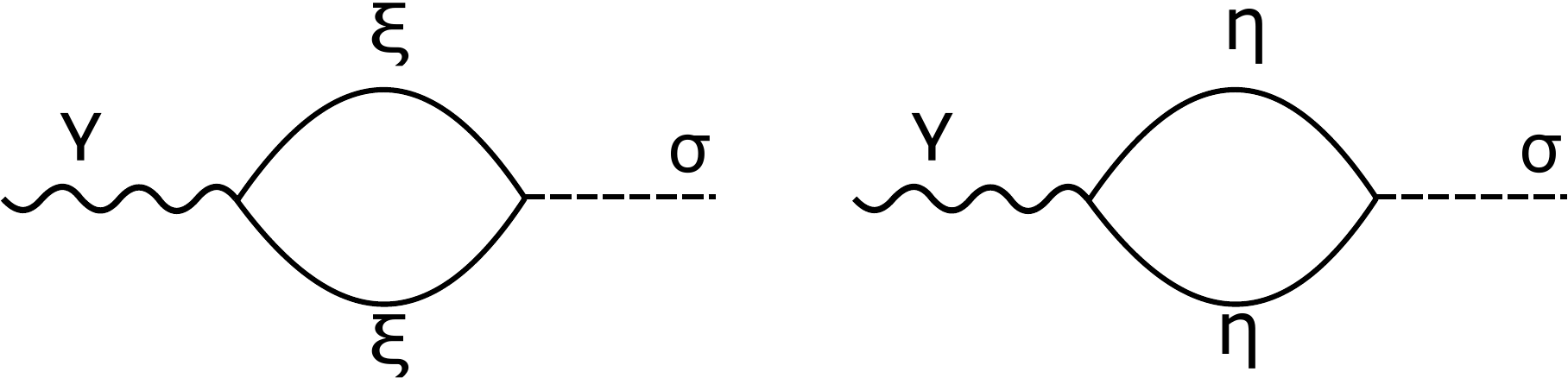}
\caption{One-loop diagrams which contribute to the the photon-scalar anomalous mixing.}
\label{fig:PhotonScalarMixingWeighted}
\end{center}
\end{figure}

The one-loop effective Lagrangians for the $\WCP{N_F-1}$ $(0,2)$ sigma-model reads  
\[\label{eq:OneLoopEffectiveAction}
\lagr = -\frac{1}{4e_\gamma^2}F_{\mu\nu}^2 + \frac{1}{e_{\sigma\,1}^2}(\dpod{\mu}\mathfrak{Re}\,\sigma)^2 + \frac{1}{e_{\sigma\,2}^2}(\dpod{\mu}\mathfrak{Im}\,\sigma)^2 +i\mathfrak{Im}(\bar{b}\,\delta\sigma)\epsilon_{\mu\nu}F^{\mu\nu}-V_{\text{eff}}(\sigma)+\text{Fermions}\,.
\]
We shall only consider photon-scalar mixing in this section, that is why we specified only bosonic part of the action.  In the above expression we denote $\sigma = \sigma_0+\delta\sigma$, where $\sigma_0$ is the VEV of the field $\sigma$ in the vacuum where our effective theory lives. 
In \eqref{eq:OneLoopEffectiveAction} effective potential $V_{\text{eff}}(\sigma)$ is given by \eqref{eq:EffectivePotentialFull},  gauge and scalar couplings can be calculated from the corresponding one-loop Feynman diagrams. Gauge field is coupled to the imaginary part of $\sigma$ and the mixing can straightforwardly be generalized from \cite{Bolokhov:2010hv}. In \figref{fig:PhotonScalarMixingWeighted} one-loop diagrams which contribute to the mixing are shown. 
The result is given by
\<
b\eq \frac{N}{4\pi}\left(\frac{1}{N}\sum\limits_{i=1}^{N-1}\frac{1}{\bar{\sigma}_0-\bar{m}_i}-\alpha\frac{1}{\tilde N}\sum\limits_{i=1}^{\tilde N-1}\frac{1}{\bar{\sigma}_0-\bar{\mu}_i}\right)\nln
\eq \frac{N}{4\pi}\frac{1}{\bar{\sigma}_0}\left(f\left(\frac{m}{|\sigma_0|}\right)-\alpha f\left(\frac{\mu}{|\sigma_0|}\right)\right)\,,
\>
where the function $f(\beta)$ was introduced in \eqref{eq:fofbeta} and we assumed that $\sigma_0\neq 0$. If the VEV for $\sigma$ vanishes at a vacuum (which happens in the symmetric \textbf{C}s phase) then the result is different
\[\label{eq:bsigma0}
b= \frac{1}{4\pi}\left(-\frac{1}{m}+\frac{\alpha}{\mu}\right)\,.
\]
The photon mass can be obtained by diagonalization of the mass Lagrangian
\[\label{eq:PhotonMassGeneric}
m_\gamma= e_{\sigma\,2} e_\gamma |b|\,.
\]
We can immediately see from \eqref{eq:bsigma0}, \eqref{eq:PhotonMassGeneric} and the formulae for the couplings \eqref{eq:PhotonCoupling}, \eqref{eq:ScalarEffCoupl}  that in the symmetric \textbf{C}s phase photon is massless in the large-$N$ approximation. This result is universal, it is dictated by the unbroken discrete $\Integers_{N-\tilde N}$ symmetry present in the \textbf{C}s phase, and it is independent of the value of the heterotic deformation.

Let us now calculate the photon case for zero and nonzero values of $u$ in the strongly coupled Coulomb phases \textbf{C}$m$ and \textbf{C}$\mu$, where discrete symmetries are spontaneously broken by the VEVs of $\sigma$.

\subsection{Undeformed $(2,2)$ Model}

If the $(2,2)$ supersymmetry is unbroken the masses of the particles of the same multiplet should be the same 
\[\label{eq:MassesUndeformed}
m_\gamma=m_\sigma=m_{\text{fermi}}\,.
\]
Using \eqref{eq:PhotonMassGeneric, eq:CounlingUndeformed} we can easily find
\[\label{eq:MassFullAnswer}
m_\gamma =\displaystyle\frac{A}{\displaystyle\frac{|\sigma_0|}{\left||\sigma_0|^2-m^2\right|}+\alpha\displaystyle\frac{|\sigma_0|}{\left||\sigma_0|^2-\mu^2\right|}}\,,
\]
where the numerator reads
\[\label{eq:ANumerator}
A=\left\vert f\left(\frac{m}{|\sigma_0|}\right)-\alpha f\left(\frac{\mu}{|\sigma_0|}\right)\right\vert\,.
\]
Depending on the VEV $\sigma_0$ the masses \eqref{eq:MassesUndeformed} can have different values, in particular, they can vanish.

\paragraph{$\Lambda$-Vacua.}

The $\Lambda$-vacua \eqref{eq:CCVacua} appear only in the \textbf{C}$m$ region. The mass of the $\ssN=2$ multiplet is given by \eqref{eq:MassFullAnswer} with $|\sigma_0|=\Lambda$ and the following numerator 
\[
A =2(1-\alpha)\,.
\]
%

\paragraph{$0$-Vacua.}

In the Coulomb phase there are also $0$-vacua which are the solutions of the vacua equations in the two regions of the parameter space \textbf{C}$m$ and \textbf{C$\mu$} (see \figref{fig:PhaseDiagram22}). In this case in the formulae \eqref{eq:MassFullAnswer, eq:ANumerator} we should use
we have
\<
A\eq 2\alpha\,, \quad |\sigma_0|=\Lambda\left(\frac{m}{\Lambda}\right)^{1/\alpha} \qquad  \text{in \textbf{C}$m$ phase}\nln
A\eq 2\,, \quad |\sigma_0|=\Lambda\left(\frac{\mu}{\Lambda}\right)^\alpha \qquad \text{in \textbf{C}}\mu\,\, \text{phase}\,.
\>
%

\subsection{Deformed $(0,2)$ Model}

As we have observed in the previous sections, $\Lambda$ vacua become metastable as we increase $u$ and for $u>u_{\text{crit}}$ disappear completely. Keeping this in mind let us focus on $0$-vacua, which continue to exist for any value of the deformation, assuming that $u$ is large enough for the approximations we have used in the end of \secref{Sec:Large-NSolution} to be valid.

In the \textbf{C}$m$ phase we get
\<
\frac{4\pi}{N e_\gamma^2}\eq \frac{1}{m^2}+\frac{\alpha}{3}\frac{1}{\Lambda^2}\left(\frac{\Lambda}{m}\right)^{2/\alpha}+\frac{2 \alpha}{3}\frac{1}{\Lambda^2\left(\frac{m}{\Lambda}\right)^{2/\alpha } e^{-\frac{u}{\alpha}}-\mu^2}\,,\nln
\frac{4\pi}{N e^2_{\sigma\,2}}\eq \frac{1}{m^2}+\frac{\alpha }{\Lambda^2\left(\frac{m}{\Lambda}\right)^{2/\alpha } e^{-\frac{u}{\alpha}}-\mu^2}\,,
\>
where we have neglected all the terms as in the calculation of VEV $D$ and $\sigma_0$. The photon mass by means of \eqref{eq:PhotonMassGeneric} is then given by
\[
m_\gamma = \sqrt{6}\, \Lambda\left(\frac{\Lambda}{m}\right)^{1/\alpha }\left(\left(\frac{m}{\Lambda}\right)^{2/\alpha }-\left(\frac{\mu}{\Lambda}\right)^2 e^{u/\alpha }\right) \mathrm{e}^{-\frac{u}{2\alpha }} \,,
\]
where we have used \eqref{eq:ANumerator} which implies that $A=\alpha$ in the \textbf{C}$m$ phase. The above expression may seem to diverge at large $u$, but we do not need to forget that the expression in the parentheses above should be bigger that zero for all $u$. The bigger $u$ is the smaller is $\mu$ and the whole expression becomes suppressed. 

Analogously, the photon mass in the \textbf{C}$\mu$ phase reads
\[
m_\gamma=\sqrt{6}\Lambda\left(\frac{\Lambda}{\mu}\right)^\alpha\left(\left(\frac{\mu}{\Lambda}\right)^{2\alpha}-\left(\frac{\mu}{\Lambda}\right)^2 e^u\right)e^{-u/2} \,,
\]
where we used that $A=2$.

\section{Conclusions and Discussion}\label{Sec:Discussion}

In this paper we solved, in the large-$N$ approximation, a particular kind of two-dimensional $\ssN=(0,2)$ non-linear sigma-model which we referred to as ``heterotic'' $\WCP{N_{F}}$. As it was already noticed, we didn't study the most general kind of heterotic deformations, rather, we focused on the particular case relevant to the study of non-Abelian vortices. The main result of the paper is the determination of the phase diagram of the theory summarized in \figref{fig:PhaseDiagLargeu}, which generalize the well-known $\ssN=(2,2)$ case (see \figref{fig:PhaseDiagram22}), once a heterotic deformation is turned on. In addition to the two Coulomb phases \textbf{C}$m$, \textbf{C}$\mu$ and the two Higgs phases \textbf{H}$n$ and \textbf{H}$\rho$, already present at zero values of the deformation parameter, a new \textbf{C}$s$  \cite{Bolokhov:2010hv} phase emerges around what we called the super-conformal line: $\mu/\Lambda=\left(m/\Lambda\right)^{1/\alpha}$. {On this line some excitations become massless, and the theory is described by a super-conformal theory of the minimal $A_{N-1}$ type \cite{Tong:2006pa,Martinec:1988zu,Vafa:1988uu,Witten:1993jg}.} A discrete $\mathbb Z_{N-\Nt}$ symmetry is broken in all phases but it is preserved in the  \textbf{C}$s$ phase. Supersymmetry is also generically broken. The vacuum energy and the expectation value of the auxiliary field $D$ vanish as we approach the super-conformal line suggesting that supersymmetry is unbroken on the line. The \textbf{C}$\mu$ phase contains two well-defined sets of vacua which we called $\mu$-vacua and $\Lambda$-vacua (in this region $\mu<\Lambda$). Once the heterotic deformation is turned on, the $\Lambda$ vacua become metastable. For sufficiently large values of the deformation ($u>u_{crit}$), the $\Lambda$ vacua do not exist at all. All the phase transitions look like being of the second order \cite{Bolokhov:2010hv}, but it is important to stress that this is an effect of the leading order large-$N$ approximation: at finite $N$, they should rather look like sharp crossovers. 

{The vacuum diagram in  \figref{fig:PhaseDiagLargeu} also gives us the spectrum of non-Abelian vortices in the associated $\ssN=1$ four-dimensional gauge theory. In particular, supersymmetry breaking means that vortices are not BPS saturated at the quantum level. Furthermore, the vacuum energy (see \eqref{eq:smallenergy}) is translated into  a correction to the classical formula $T=2 \pi \xi$ for the tension of vortices\footnote{$\xi$ is the four-dimensional Fayet-Iliopoulos}. In the \textbf{C}$m$ regime, for example, $N-\Nt$ vortices become metastable and eventually disappear from the spectrum. Moreover, as was shown in Ref. \cite{Witten:1978bc} and later discussed in Refs. \cite{Hanany:1997vm,Gorsky:2005ac,Bolokhov:2009wv,Bolokhov:2010hv},  the fundamental fields $n$'s and $\rho$'s, together with their fermionic superpartners, can be interpreted as kinks interpolating between two vacua, or vortices.  As already mentioned, kinks correspond to monopoles in the four-dimensional gauge theory. The study of spectrum of the model we considered, thus, gives informations about the monopole spectrum in $\ssN=1$ theories.} 

The discovery that non-Abelian vortices are the precise link between  two and four-dimensions is a recent exciting result in the study  of supersymmetric gauge theories. Given the tighten relationship between  two-dimensional $\ssN=(2,2)$ and four-dimensional $\ssN=2$ theories, it is tempting to explore systems with less supersymmetry,  to find if and how the physics of $\ssN=1$ theories is ``seen'' on the two-dimensional $\ssN=(0,2)$ theory, and vice versa. Interesting results have already been  found when no additional hypermultiplets have been considered \cite{Tong:2007qj,Bolognesi:2009ye} where a qualitative matching of the supersymmetry braking pattern and of the meson spectrum was observed. {Our results are the first important step to extend this line of research when an additional number of flavors $\Nt$ is included. In this case,   physics of $\ssN=1$ SQCD varies dramatically \cite{Seiberg:1994pq,Argyres:1996eh}.} The most remarkable feature is the existence of an electric-magnetic duality (Seiberg duality). More recently it was found that dynamical supersymmetry breaking is a quite common feature \cite{Intriligator:2006dd}. It would be interesting to search for signs of these phenomena on the two-dimensional side. {With the better understanding of the quantum physics of vortices when additional flavors are included, it should be possible to extend, for example, the analysis made in  Refs. \cite{Gorsky:2007ip,Shifman:2007kd,Bolokhov:2009mv}, where the role of vortices in the context of Seiberg duality was investigated. In these works, the dual quarks of the ``electric'' theory were interpreted as monopoles of the ``magnetic'' theory.}

The investigations on ``heterotic'' $\WCP{N_F-1}$ are by no means finished here. First of all we think it may be interesting to further study the model on the super-conformal line. This line as a direct counterpart in the four-dimensional gauge theory, where the coalescence of multiple vacua give rise to the appearance of super-conformal vacua called Argyres-Douglas points, where the relevant degrees of freedom are mutually non-local \cite{Argyres:1995jj,Argyres:1995xn} . An analysis of this kind has been initiated for example in Ref. \cite{Tong:2006pa}. A comprehensive study of kinks is also in order. The spectrum of kinks is in fact related to the monopole spectrum in four dimensions. This study started in Ref. \cite{Shifman:2010id}. A careful study of kinks interpolating between different kind of vacua is still to be done.

\section*{Acknowledgments}

We would like to thank M. Shifman, A. Vainshtein, and M. Voloshin for many useful discussions and the preliminary reading of the paper. This work was supported in part by the DOE grant DE-FG02-94ER40823 (P.K., W.V.); by the Stanwood Johnston grant from the Graduate School of University of Minnesota, RFBR Grant No. 07-02-00878 and by the Scientific School Grant No. NSh-3036.2008.2. (A.M.). P.K. would like to thank the organizers of ESI Programme on AdS Holography and the Quark-Gluon Plasma in Vienna, October 2010, where a part of his work was done.

\appendix

\section{Superfield Formalism}\label{Sec:SuperField}

In this section we present the superfield derivation of the Lagrangian \eqref{eq:complagr} of the weighted sigma model without twisted masses.  Inclusion of twisted masses \eqref{eq:WCPNTwistedMasses} can naturally be realized in the brane picture. Here we briefly review the derivation of the Lagrangian given in Ref. \cite{Koroteev:2010gt}
\<
\label{eq:N1Heterotic}
\lagr_{W\mathbb{CP}^N}^{het}   \eq   \displaystyle\int d^2 \theta\, \Bigg[\sfrac{1}{4}\varepsilon_{\beta\alpha}(\covder_\alpha+i\mathcal{A}_\alpha){\vecs}^\dag_ i (\covder_\beta-i\mathcal{A}_\beta)\vecs_i +\sfrac{1}{4}\varepsilon_{\beta\alpha}(\covder_\alpha- i\mathcal{A}_\alpha){\mathcal{R}}^\dag_ j (\covder_\beta+i\mathcal{A}_\beta)\mathcal{R}_j + \nl
+
i \const\left(\sum\limits_{i=1}^ {N } \vecs_i^\dag \vecs_i-\sum\limits_{j=1}^{\Nt} \mathcal{R}_j^\dag \mathcal{R}_j-r _ 0\right)  \nl 
 + \sfrac{1}{4}\varepsilon_{\beta \alpha}\covder_\alpha\Bfield^\dag\,\covder_\beta\Bfield
+\l ( i \, \omega \, \Bfield (\const-\sfrac{i}{2}\bar{\covder}\gamma^5\mathcal{A}) + \text{H.c.} \r ) \Bigg]\,,
\>
where the covariant derivative is given in \eqref{eq:CovDer} and in the third line of the above Lagrangian it is implied that
\[
\overline{\covder}\gamma^5\mathcal{A} = \covder _ \a \l ( \g ^ 0 \gamma^5 \r ) _ {\a \b} \mathcal{A} _ \b\,,
\]
isovector superfields are represented by
\<\label{eq:IsoSuperField}
\vecs^i \eq n^i + \bar{\theta}\xi ^ i + \half\bar{\theta}\theta F^i, \quad i=1,\dots,N, \nln
\mathcal{R}^j \eq \rho^j + \bar{\theta}\eta ^ j + \half\bar{\theta}\theta G^j, \quad j =  1,\dots, \Nt\,,
\>
constraint superfield 
\[\label{eq:ConstrSuperField}
\mathcal{S} = \s _ 1 + \bar \t u + \half \bar \t \t D
\]
gets multiplied by the D-term constraint in the second line of \eqref{eq:N1Heterotic}, and spinor superfield under the proper gauge\footnote{see Ref. \cite{Koroteev:2010gt} for further details} 
\[\label{eq:SpinorSuperField}
\mathcal{A}_\alpha = -i (\gamma^\mu \theta)_\alpha A_\mu + (\gamma^5\theta)_\alpha \s _ 2 
+\bar{\theta}\theta\, v _ \a\nln\,.
\]
The heterotic deformation is conducted by the chiral  field
\[\label{eq:HeteroticDef}
\Bfield = - \bar \t \z + \half \bar \t \t \bar {\mathcal {F}} \mathcal {F}.
\]
which by definition contains only the right-handed fermion:
\[
\zeta=
\left(
\begin{array}{c}
  \zeta_R   \\
   0 
\end{array}
\right)
\]

In the formulae \eqref{eq:N1Heterotic}-\eqref{eq:HeteroticDef} $\theta$ is a Majorana spinor, $\sigma_1$, $\sigma_2$, $A_\mu$, $u_\alpha$, $v_\alpha$ and $D$ are real fields, while $\zeta$ and $\mathcal{F}$ are complex fields. The complex-valued parameter $\omega$ stands for the heterotic deformation. The complex-valued fields $\sigma$ and $\lambda$ from 
\eqref{eq:ConstrSuperField, eq:SpinorSuperField} can be assembled using the components of $\mathcal{S}$ and $\mathcal{A}_\alpha$ as follows
\[
\sigma = \sigma_1+i\sigma_2\,,\quad \lambda_\alpha = u_\alpha+ i v_\alpha\,.
\]
%

\section{One-loop Effective Action}\label{Sec:SomeFormulae}

\begin{figure}[ht]
\begin{center}
\includegraphics[height=5cm, width=10.5cm]{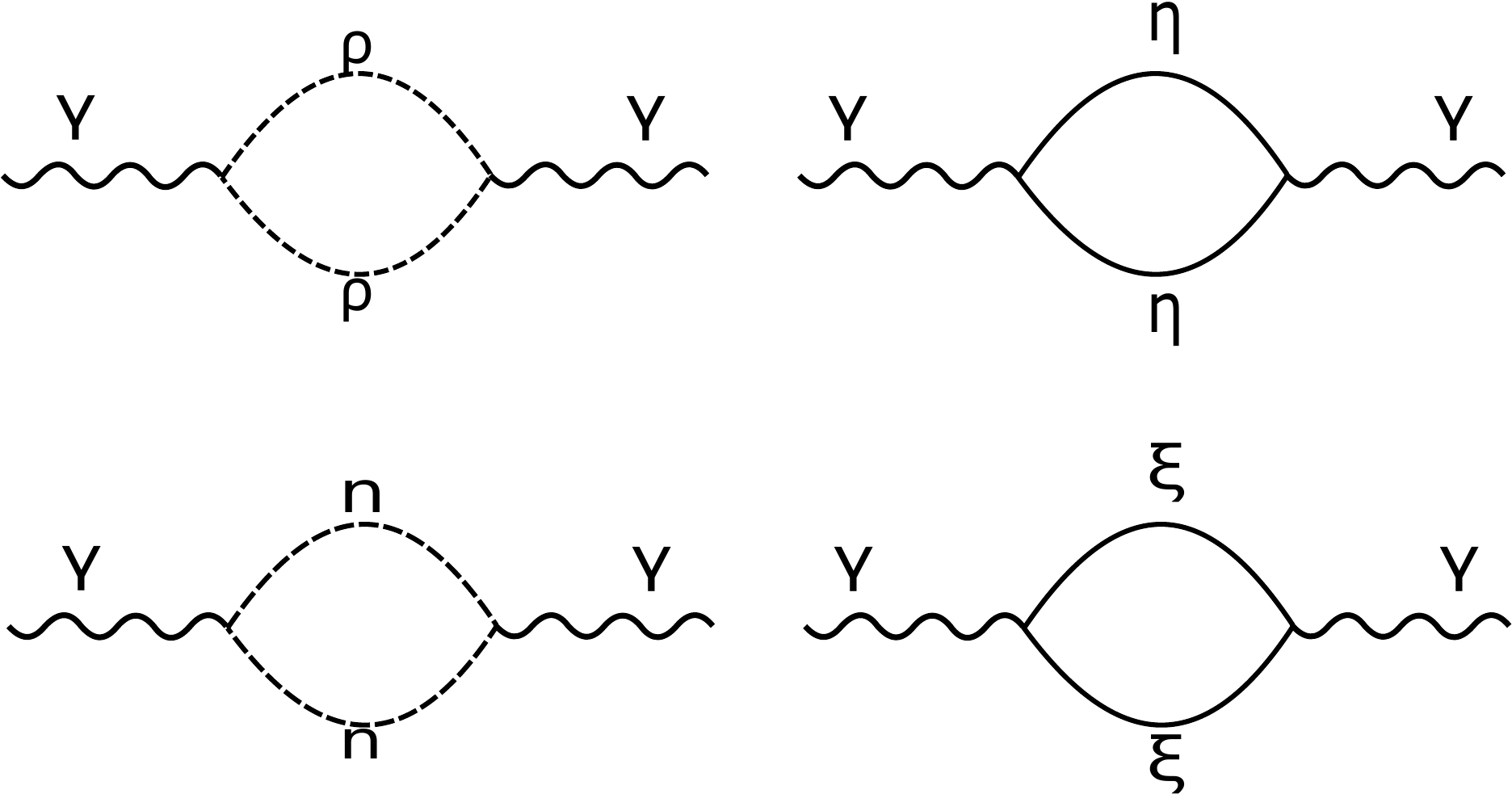}
\caption{Four series of one-loop diagrams which give photon wavefunction renormalization.}
\label{fig:PhotonWaveRenWeigted}
\end{center}
\end{figure}

\begin{figure}[ht]
\begin{center}
\includegraphics[height=5cm, width=10.5cm]{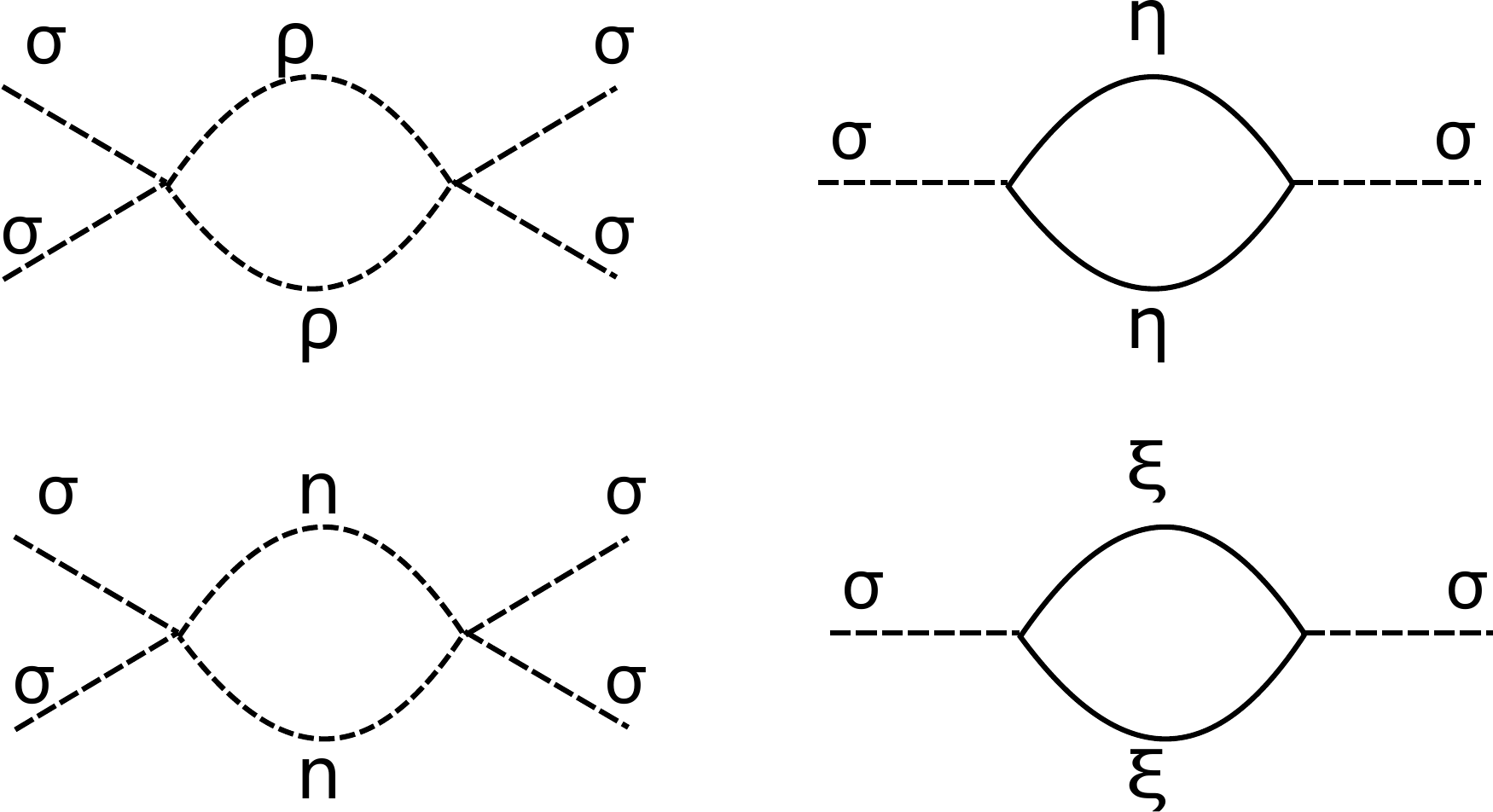}
\caption{Four series of one-loop diagrams which give scalar wavefunction renormalization.}
\label{fig:ScalarCouplingRenWeigted}
\end{center}
\end{figure}

\begin{figure}[ht]
\begin{center}
\includegraphics[height=2.4cm, width=10.5cm]{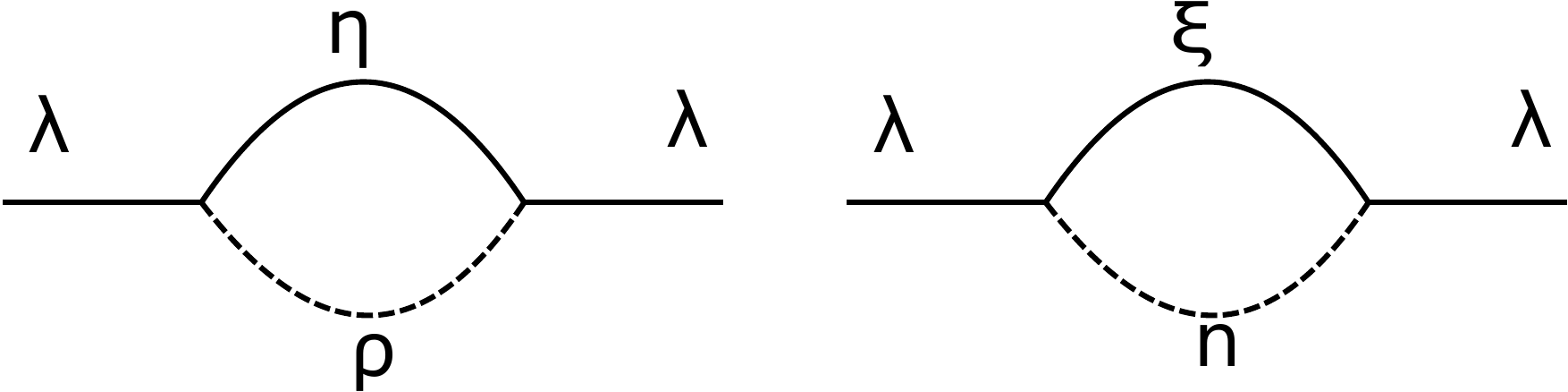}
\caption{Two series of one-loop diagrams which give scalar wavefunction renormalization.}
\label{fig:LambdaCouplingRenWeigted}
\end{center}
\end{figure}

Below we list generic expressions for the effective couplings of \eqref{eq:OneLoopEffectiveAction} in terms of  $D,\sigma, u$ and twisted masses. Let us first for completeness specify the full action including the fermionic part
\<
\lagr \eq -\frac{1}{4e_\gamma^2}F_{\mu\nu}^2 + \frac{1}{e_{\sigma\,1}^2}(\dpod{\mu}\mathfrak{Re}\,\sigma)^2 + \frac{1}{e_{\sigma\,2}^2}(\dpod{\mu}\mathfrak{Im}\,\sigma)^2 + i\frac{1}{e_\lambda^2}\bar{\lambda}\gamma^\mu\nabla_\mu \lambda+\frac{i}{2}\bar{\zeta}_R \dpod{L}\zeta_R \nl
+i\mathfrak{Im}(\bar{b}\,\sigma)\epsilon_{\mu\nu}F^{\mu\nu}   -V_{\text{eff}}(\sigma)-(i\Gamma \bar{\sigma}\bar{\lambda}\lambda+i\omega\lambda_L\zeta_R+\text{H.c.})\,.
\>
The gauge coupling can be calculated using the wavefunction renormalization for the photon \figref{fig:PhotonWaveRenWeigted} 
and reads\footnote{We assume for simplicity that for the vacuum $\mathfrak{Im}\, \sigma_0=0$.}
\<\label{eq:PhotonCoupling}
\frac{1}{e_\gamma^2} \eq\frac{1}{4\pi}\sum\limits_{i=1}^{N-1}\left[\frac{1}{3}\frac{1}{|\sigma_0-m_i|^2+D}+\frac{2}{3}\frac{1}{|\sigma_0-m_i|^2}\right]\nl
+\frac{1}{4\pi}\sum\limits_{i=1}^{\tilde N-1}\left[\frac{1}{3}\frac{1}{|\sigma_0-\mu_i|^2-D}+\frac{2}{3}\frac{1}{|\sigma_0-\mu_i|^2}\right]\,.
\>
Feynman diagrams corresponding to the scalar couplings renormalization can be found in \figref{fig:ScalarCouplingRenWeigted}. 
Performing the integrals we obtain
\<\label{eq:ScalarEffCoupl}
\frac{1}{e_{\sigma\,1}^2}\eq \frac{1}{4\pi}\sum\limits_{i=1}^{N -1}\frac{1}{|\sigma_0-m_i|^2}\left[\frac{1}{3}+\frac{2}{3}\frac{|\sigma_0-m_i|^4}{(|\sigma_0-m_i|^2+D)^2}\right] +\frac{1}{4\pi}\sum\limits_{i=1}^{\tilde N -1}\frac{1}{|\sigma_0-\mu_i|^2}\left[\frac{1}{3}+\frac{2}{3}\frac{|\sigma_0-\mu_i|^4}{(|\sigma_0-\mu_i|^2-D)^2}\right]\,,\nln
\frac{1}{e_{\sigma\,2}^2}\eq \frac{1}{4\pi}\sum\limits_{i=1}^{N -1}\frac{1}{|\sigma_0-m_i|^2}+\frac{1}{4\pi}\sum\limits_{i=1}^{\tilde N -1}\frac{1}{|\sigma_0-\mu_i|^2}\,.
\>
We see that real and imaginary components of the $\sigma_0$ field acquire different renormalizations, in particular, is the SUSY is broken, their couplings are different.  The fermion coupling renormalization given by the diagrams in \figref{fig:LambdaCouplingRenWeigted}
reads
\<\label{eq:FermionEffectiveCoupl}
\frac{1}{e_\lambda^2}\eq \frac{N-\tilde N}{4\pi}\frac{2}{D}-
\frac{1}{2\pi}\sum\limits_{i=1}^{N-1}\frac{|\sigma_0-m_i|^2}{D^2}\log\frac{|\sigma_0-m_i|^2+D}{|\sigma_0-m_i|^2}\nl
-\frac{1}{2\pi}\sum\limits_{i=1}^{\tilde N-1}\frac{|\sigma_0-\mu_i|^2}{D^2}\log\frac{|\sigma_0-\mu_i|^2-D}{|\sigma_0-\mu_i|^2}\,.
\>
The Yukawa coupling can be found as the mass renormalization using \figref{fig:LambdaCouplingRenWeigted} and is given by (equivalently one could compute the corresponding triangular graph)
\[
\Gamma=\frac{1}{4\pi}\frac{2}{D}\left[\sum\limits_{i=1}^{N-1}\log\frac{|\sigma_0-m_i|^2+D}{|\sigma_0-m_i|^2}+\sum\limits_{i=1}^{\tilde N-1}\log\frac{|\sigma_0-\mu_i|^2-D}{|\sigma_0-\mu_i|^2}\right]\,.
\]

The sums in the above formulae can be done explicitly. Below we list those formulae we used in \secref{Sec:Spectrum}.  Using
\[\label{eq:SumAbsSq}
\frac{1}{N}\sum\limits_{k=1}^{N-1}\frac{1}{|1-\gamma e^{\frac{2\pi i k}{N}}|^2}=\frac{1}{|1-\gamma^2|}\,,
\]
and, for nonzero $D$
\[\label{eq:SumGammaCos}
\frac{1}{N}\sum\limits_{k=1}^{N-1}\frac{1}{1-\gamma\cos{\frac{2\pi i k}{N}}}=\frac{1}{\sqrt{1-\gamma^2}}\,,
\]
the value of the gauge coupling \eqref{eq:PhotonCoupling} can be evaluated and reads
\<
\frac{1}{e_\gamma^2}\eq \frac{N}{4\pi}\left[\frac{1}{3}\frac{1}{\sqrt{\left(|\sigma_0|^2+m^2+D\right)^2-4|\sigma_0|^2m^2}}+\frac{2}{3}\frac{1}{\left\vert |\sigma_0|^2-m^2\right\vert}\right]\nl
+\frac{\tilde N}{4\pi}\left[\frac{1}{3}\frac{1}{\sqrt{\left(|\sigma_0|^2+\mu^2-D\right)^2-4|\sigma_0|^2\mu^2}}+\frac{2}{3}\frac{1}{\left\vert |\sigma_0|^2-\mu^2\right\vert}\right]\,,
\>
whereas the coupling for the imaginary part of $\sigma$ is given by
\[
\frac{1}{e_{\sigma\, 2}^2}= \frac{N}{4\pi}\frac{1}{\left\vert |\sigma_0|^2-m^2\right\vert}+\frac{\tilde N}{4\pi}\frac{1}{\left\vert |\sigma_0|^2-\mu^2\right\vert}\,.
\]
Thus if the SUSY is unbroken we can get from the above two formulae and \eqref{eq:FermionEffectiveCoupl} the following
\[\label{eq:CounlingUndeformed}
\frac{1}{e_\gamma^2}=\frac{1}{e_{\sigma\,1}^2}=\frac{1}{e_{\sigma\,2}^2}=\frac{1}{e_\lambda^2}=\frac{N}{4\pi}\left(\frac{1}{\left||\sigma_0|^2-m^2\right|}+\alpha\frac{1}{\left||\sigma_0|^2-\mu^2\right|}\right)\,.
\]
%

\section{Notations}\label{Sec:Notations}

Here we list the notations and some useful relations we use in the paper.

Gamma matrices
\[
\gamma^0 =  \sigma_2 =
\begin{pmatrix}
0& -i\\
i&0
\end{pmatrix}\,, \quad
\gamma^1 =  i\sigma_1 =
\begin{pmatrix}
0& i\\
i&0
\end{pmatrix}\,,\quad
\gamma^5 = \gamma^0 \gamma^1 = \sigma_3 =
\begin{pmatrix}
1& 0\\
0&-1
\end{pmatrix}\,.
\]
Antisymmetric symbol
\be
\varepsilon _ {\a \b} = \l (
\ba{cc}
0 & 1 \\
-1 & 0 
\ea
\r ).
\ee
Left and right coordinates
\[
\begin{array}[b]{rclcrclcrcl}
x_L \eq x_0 + x_1, &&
\p _ 0 \eq \p _ L + \p _ R, &&
\p _ L \eq \half \l ( \p _ 0 + \p _ 1 \r ),
\\
x_R \eq x_0 - x_1, &&
\p _ 1 \eq \p _ L - \p _ R, && 
\p _ R \eq \half \l ( \p _ 0 - \p _ 1 \r )\,.
\\
\end{array}
\]
Left and right fermions
\<
\psi \eq 
\l(
\ba{c}
\psi _ R\\
\psi _ L
\ea
\r )
\>
are eigenstates of $\gamma^5$
\[
\gamma^ 5 \psi _ {R,L} = \pm \psi _ {R,L}\,.
\]
Derivatives and integrals
\<
\int d^ 2 \t \,\bar \t \t \eq \int d \t _ 1 \, d \t _ 2 \bar \t \t = \int d \t _ 1 \, d \t _ 2 \, 2 i \t _ 2 \t _ 1 = 2 i\,, \nln
\f { \p } {\p \bar \t _ \a } \t _ \b \eq \g ^ 0 _ {\a \b}\,.
\>
Contraction of indices for Majorana fermions
\be
\bar { \psi } \t = \psi ^ \dagger \g ^ 0 \psi = \psi ^ T \g ^ 0 \psi 
=i \t _ 2 \psi _1 - i \t _ 1 \psi _2 = \bar {\theta} \psi\,,
\ee
\be
\bar {\t} \g ^ {0,1} \t = \l ( \t _ 1 \r ) ^ 2 = \l ( \t _ 2 \r )^ 2 = 0\,,
\ee
\bea
\bar {\t}  \t = 2 i \t _ 2 \t _ 1 = - 2 i \t _ 1 \t _ 2\,, \nonumber \\
\t _ \a \t _ \b = \f {i} {2} \epsilon _ {\a \b} \, \bar {\t}  \t = - \half
\g ^ 0 _ {\a \b} \bar {\t}  \t\,, \nonumber \\
\bar {\t} _ \a  \t _ \b = \half \delta _ {\a \b} \, \bar {\t}  \t\,.
\eea
Some relations for gamma matrices
\<
\g^{\m T} \eq - \g ^ 0 \g ^ \m \g ^ 0\,, \nln
\g^{\m \dagger} \eq \g ^ 0 \g ^ \m \g^0\,.
\>
%

\paragraph{Supersymmetry transformations.}

Coordinate transformations 
\bea
x _ \m & \to & x _ \m + i \bar \eps \g  _ \m \t, \nonumber \\
\t _ \a & \to & \t _ \a + \eps _ \a, \nonumber \\
\bar {\t} _ \a & \to & \bar { \t } _ \a + \bar { \eps } _ \a.
\eea
Chiral superfield 
\[\label{eq:ChiralSuperField}
\Phi = \phi+ \bar{\theta}\psi + \half \bar{\theta}\theta F\,, 
\]
obeys the following supertransformations
\<
\delta n\eq \bar {\eps} \psi, \nln
\delta \psi \eq -i \p _ \m n\g ^ \m \eps + \eps F, \nln
\delta F \eq - i \bar {\eps} \g ^ \m \p _ \m \psi\,.
\>
A natural generalization of the chiral superfield is the isovector superfield
\[
\vecs^i = n^i+ \bar{\theta}\psi^i + \half \bar{\theta}\theta F^i\,,\quad  i=1,\dots, N\,.
\]
Supertransformations act as
\[
\delta \Phi = \bar \epsilon \gen{Q} \Phi\,
\]
Supersymmetry generators
\<
\gen{Q}_\alpha \eq \f {\p} {\p \bar \t _ \a } -i \left(\gamma^\mu \theta \right)_\alpha \dpod{\mu} \,,
\>
Covariant derivative
\[
\covder_\alpha = \f { \p } {\p \bar \t _ \a } +i \l ( \g ^ \m \t \r ) _ \a \dpod{\mu}\,,
\]
anticommutes with the supercharge
\[
\{\gen{Q}_\alpha, \covder_\beta\} = 0\,.
\]
%

\paragraph{Chiral Notation.}

One can use the following identification
\[
x^\mu = \gamma^\mu_{\alpha\beta} x^{\alpha\beta}\,,\quad \mu=1,2\,, \quad \alpha,\beta =1,2\,.
\]
Having done so we can write
\[
\gen{Q}_{\alpha} =\epsilon_{\alpha\beta} \po{\theta_\beta}+\theta_\beta\dpod{\alpha\beta}\,.
\]
Accordingly we have
\<
\acomm{\gen{Q}_1}{\gen{Q}_1} \eq 2\dpod{12} = 2\left(i\po{t}-i\po{x}\right)=2(\ham+\moment)\,,\nln
\acomm{\gen{Q}_2}{\gen{Q}_2} \eq 2\dpod{21} = 2\left(i\po{t}+i\po{x}\right)=2(\ham-\moment)\,,\nln
\acomm{\gen{Q}_1}{\gen{Q}_2} \eq 0\,,
\>
where $\ham$ and $\moment$ are energy and momentum charges respectively. Covariant derivative reads
\[
\covder_{\alpha} =i\epsilon_{\alpha\beta} \po{\theta_\beta}-i\theta_\beta\dpod{\alpha\beta}\,.
\]
%


\bibliography{cpn1}
\bibliographystyle{nb}

\end{document}